\documentclass[journal]{IEEEtran}
\ifCLASSINFOpdf
\else
\fi

\usepackage{graphicx}%²åÈëÍ¼Æ¬
\usepackage{bm}%´ÖÌå×Ö°ü
\usepackage{amsfonts}
\usepackage{amsmath}
\usepackage{amssymb}
\usepackage{comment}
\usepackage{times}
\usepackage{subfigure}
\usepackage{latexsym,bm,amsmath,amssymb} %Latex ÖÐÊýÑ§°ü
\usepackage{CJK}
\usepackage{hhline}
\usepackage{mathtools}
\usepackage{multirow} %multirow for format of table
\usepackage{amsmath}
\usepackage{xcolor}
\usepackage{epstopdf}
\usepackage{cite}
\usepackage[noend]{algpseudocode}
\usepackage{algorithmicx,algorithm}
\usepackage{dblfloatfix}
\usepackage{flushend}
\usepackage{comment}
\begin{document}
\title{Faster-than-Nyquist Signaling is Good for Single-Carrier ISAC: An Analytical Study}

\author{Shuangyang Li,~\IEEEmembership{Member,~IEEE,}
Fan Liu,~\IEEEmembership{Member,~IEEE,} 
Yifeng Xiong,~\IEEEmembership{Member,~IEEE,}
Weijie Yuan,~\IEEEmembership{Member,~IEEE,}
Baoming Bai,~\IEEEmembership{Senior Member,~IEEE,} 
Christos Masouros,~\IEEEmembership{Fellow,~IEEE,}
and Giuseppe Caire,~\IEEEmembership{Fellow,~IEEE}

\thanks{
Part of this paper was submitted to IEEE Information Theory Workshop 2025~\cite{Shuangyang2025ITW}. \\
S. Li and G. Caire are with the Faculty of Electrical Engineering and Computer Science, Technical University of Berlin, Germany (e-mail:\{shuangyang.li, caire\}@tu-berlin.de).\\
F. Liu is with the National Mobile Communications Research Laboratory,
Southeast University, China (e-mail: f.liu@ieee.org).\\
Y. Xiong is with the School of Information and Electronic Engineering,
Beijing University of Posts and Telecommunications, China. (e-mail: yifengxiong@bupt.edu.cn).\\
W. Yuan is with the School of System Design and Intelligent Manufacturing, Southern University of Science and Technology,  China. (e-mail: yuanwj@sustech.edu.cn).\\
B. Bai is with the State Key Lab. of ISN, Xidian University, Xi’an 710071, China. (e-mail: bmbai@mail.xidian.edu.cn).\\
C. Masouros is with the Department of Electrical and Electronic Engineering, University College London, UK (email: c.masouros@ucl.ac.uk).
}}

\maketitle

\begin{abstract}
 In this paper, we provide an analytical study of single-carrier faster-than-Nyquist (FTN) signaling for integrated sensing and communications (ISAC). Our derivations show that FTN is advantageous for ISAC, and reveal new insights that these advantages come from the fact that FTN signaling can effectively avoid the spectral aliasing due to the mismatch between the symbol rate and the bandwidth of the shaping pulse. Specifically, the communication spectral efficiency advantages of FTN signaling over time-invariant multipath channels are analytically shown, where both upper- and lower-bounds on the spectral efficiency are derived. We show that the gap between these two bounds corresponds to the potential signal-to-noise ratio (SNR) variation due to the presence of multipath delay and spectral aliasing, which diminishes as the symbol rate grows higher. Particularly, in the limiting case, this SNR variation disappears while the degree of freedom (DoF) of the system attain the maximum.
Furthermore, the sensing advantages for FTN signals are verified in terms of the expected normalized squared ambiguity function. We show that FTN signals generally enjoy a more robust ranging performance. More importantly, we prove that FTN signaling can effectively avoid the undesired peaks in the considered ambiguity function along the Doppler dimension, thereby reducing the ambiguities in velocity estimation.
All these conclusions are explicitly verified by numerical results.
\end{abstract}


% \fi
\IEEEpeerreviewmaketitle

\section{Introduction}
Integrated sensing and communications (ISAC) is listed as one of the six usage scenarios in the 6G recommendation~\cite{ITU2023}. It is expected to play a vital role for facilitating emerging applications and services that require sensing capabilities, such as autonomous vehicles~\cite{Qixun2022TDISAC} and extended reality~\cite{EricssonBlog}.  
The key feature of ISAC is the accommodation of both communication and sensing functionalities using a single well-designed signal~\cite{Fan_ISAC_JSAC,Fan2020JRC,Zichao2024novel}, which requires the signal to be, on one hand, robust against communication channel dynamics, and on the other hand, suitable for exploiting the sensing channel characteristics. 
This particular requirement has motivated recent studies on the physical layer waveform design for pursuing favorable performance tradeoffs between communications and sensing~\cite{Wenxing2022ISAC,Fan2018optimal,Peishi2025MIMOOFDM}.

A fundamental difference between typical communication signals and sensing signals is the fact that communication signals need to modulate \textit{random} information symbols, while the sensing signal usually prefers a \textit{deterministic} pattern that is well-optimized for sensing parameter exploitation, e.g., chirp signals~\cite{cook2012radar}. 
This subtle difference leads to the recent studies on the fundamental tradeoff between communications and sensing in~\cite{Yifeng2023fundamental}, where the authors quantitatively evaluated the significance of the deterministic-random tradeoff (DRT) in point-to-point ISAC over Gaussian channels. 
Although the solution for achieving the optimal communication and sensing tradeoff is derived in~\cite{Yifeng2023fundamental}, the required DRT could be hard to achieve in practice, especially in next-generation wireless networks. This is because of the exceedingly high data rate required in future wireless networks, which inevitably pushes the ISAC transmission to a communication-centric mode~\cite{Fan2020JRC}.
Therefore, an interesting and practically important study is to characterize the sensing performance using random communication signals, which is the motivation of this work. 

The sensing performance under random communication signals has been evaluated in several recent studies. In~\cite{Zhen2024TSP}, the sensing performance using orthogonal frequency-division multiplexing (OFDM) communication signals has been studied. 
Particularly, an optimal probabilistic constellation shaping approach was proposed to maximize the achievable information rate under the power constraint, which is developed based on a modified Blahut-Arimoto algorithm~\cite{yeung2008information}.
The optimality of OFDM for achieving the lowest ranging sidelobe using random communication signals with practical constellations has been verified in~\cite{liu2024ofdm}. Interestingly,~\cite{liu2024ofdm} also reveals that the sidelobes for ranging and Doppler sensing using random communication signals have a close connection to the \textit{kurtosis} of the constellation alphabet from which the information symbols are randomly taking values. 
In this context, the Gaussian constellation was proved to be a special case, whose sensing performance is independent of the choice of the communication modulation format. Furthermore, ``sub-Gaussian'' and ``super-Gaussian'' constellations are defined according to the values of their kurtosis, whose optimal ranging performances are proved to be achieved by OFDM and single-carrier transmissions, respectively. 
The importance of pulse shaping for sensing using random communication signals is studied in~\cite{liu2024iceberg}, where the connections between the shaping pulse and the data sequence on the expected squared ambiguity function for typical Nyquist communication signals were highlighted. Furthermore, the pulse design for sensing using single-carrier communication signals was considered in~\cite{liao2024pulseshapingrandomisac}. The 
pulse optimization issue is formulated subjected to the constraints on 
the Nyquist no intersymbol interference (ISI) condition, power constraint, and a reasonable out-of-band emission, which is then solved by using successive convex approximation (SCA) and alternating direction method of multipliers (ADMM) approaches. The resultant designed pulse yields a significant sensing performance improvement compared with sensing using root-raised cosine (RRC) pulse shaped communication signals. All the aforementioned works were focused on Nyquist communication signals, while the performance of sensing using non-Nyquist  communication signals, to the best knowledge of the authors, remains unexplored. 

Against the above background, we study the impact of non-Nyquist signaling for ISAC under the framework of faster-than-Nyquist (FTN) signaling in this paper. FTN signal is a typical non-Nyquist (non-orthogonal) signal that was firstly considered by Shannon in the landmark paper~\cite{Shannon1948mathematical} and then popularized by Mazo in~\cite{FTNMAZO}, which intentionally introduces controllable ISI at the transmitter side in order to improve the communication efficiency~\cite{FTNANDERSON}.
From a practical point of view, FTN signaling is an important bandwidth-efficient waveform suitable to meet the stringent data rate requirements of wireless networks. 
The application of FTN signaling in satellite communications was considered in~\cite{Piemontese2013TFP}, where a significant boost of spectral efficiency can be observed by using a practical iterative detection and decoding scheme. The suitability of FTN signaling in 5G network was studied in~\cite{Paolo2014Heir}, where the authors discussed the potential advantages of single-carrier FTN signaling for uplink communications in the presence of large antenna arrays. One of the reasons for this suitability is the low peak-to-average power ratio (PAPR) property of single-carrier FTN signaling, which was also highlighted in~\cite{Chung2014benefit}.
The practical application of FTN signaling in achieving high data rate was considered in~\cite{Ibrahim2021OOJCS}, where a practical detector capable of supporting FTN with high-order quadrature-amplitude modulation (QAM) constellation (up to $65536$-QAM) was proposed. The proposed detector was developed according to the ADMM approach and exhibited a good tradeoff between the performance and complexity. 
The hardware implementation of FTN signaling based on CMOS and FPGA was reported in~\cite{Dasalukunte2011multicarrier}, which demonstrates the fact that FTN signaling can be used to achieve higher bandwidth efficiency with acceptable complexity overhead.
A dedicated chip designed for FTN signaling was reported in~\cite{dasalukunte2014faster}, which achieves a twofold improvement in bandwidth usage with similar performance as that of an OFDM system. 
Furthermore, the FTN prototype in narrow-band internet-of-things (IoT) application was reported in~\cite{Tongyang2019wavefrom}, where a roughly $11 \%$ improvements in data rate was demonstrated.
From a theoretical point of view, the application of ISAC using the FTN signal is inherently interesting due to its unique spectral properties. In the conventional Nyquist transmission, the symbol rates are often mismatched with the bandwidth of the practical Nyquist pulses, such as RRC pulses, since the perfect sinc pulse is not realizable in practice~\cite{FTNANDERSON}. Consequently, the spectrum of Nyquist communication signals usually suffers from \textit{spectral aliasing} due to the excess bandwidth required for achieving orthogonality. On the contrary, this aliasing can be effectively avoided by simply transmitting symbols faster. As a result, FTN signaling is capable of exploiting the degrees of freedom (DoFs) offered by the excess bandwidth of the shaping pulse~\cite{rusek2009constrained} for enhancing the communication spectral efficiency. This improvement in spectral efficiency has been witnessed in many scenarios, including but not limited to additive white Gaussian noise (AWGN) channels~\cite{rusek2009constrained}, Gaussian broadcast channels~\cite{kim2016faster}, and multiple-input multiple-output (MIMO) channels
~\cite{Rusek2007MIMOFTN,Zichao2023FTN}.
More importantly, it is not clear in the literature how the spectral aliasing will affect the sensing performance, since the typical radar waveform does not need to modulate random information symbols and therefore the spectral aliasing will not appear in general.

In this paper, we study the communication and sensing performance of single-input single-output (SISO) single-carrier FTN signaling, where the impact of spectral aliasing is  thoroughly considered. Specifically, we derive the spectral efficiency of single-carrier FTN over time-invariant multipath channels. Different from~\cite{Ishihara2021icc}, which assumes that the channel state information (CSI) is known at the transmitter, we consider the practical case where the CSI is available only at the receiver and therefore, the optimal water-filling solution is not applicable{\footnote{It should be noted that the capacity-optimal precoded FTN signal is equivalent to a Nyquist signal shaped by a sinc pulse with a larger effective bandwidth, as implied by~\cite{rusek2009constrained}. }}. Furthermore, the sensing performance of FTN signals is evaluated by considering the expected squared ambiguity function that is a general performance metric independent from the underlying sensing channel condition. 
The major contributions of this paper are summarized as follows.
\begin{itemize}
    \item We first evaluate the effective communication channel matrix of FTN transmissions and verify that it is asymptotically Hermitian Toeplitz. Thanks to this property, we are able to derive the communication spectral efficiency based on the Szeg{\"o}'s theorem, which relies on the discrete-time Fourier transform (DTFT) of the Toeplitz coefficients. 
    \item We derive the upper- and lower-bounds on the Toeplitz coefficients' DTFTs, which explicitly take into account the impact of spectral aliasing and multipath delay. Relying on these bounds, we further derive the upper- and lower-bounds on the spectral efficiency of FTN signaling over time-invariant multipath channels. Our analysis shows that the channel delay leads to potential signal-to-noise ratio (SNR) variation for FTN signaling via the spectral aliasing and this SNR variation reduces as the symbol rate grows higher. Particularly, in the limiting case, where the symbol rate is no smaller than the signal bandwidth, both derived upper- and lower-bounds on the spectral efficiency coincide with each other and the DoF attains its maximum while the SNR variation disappears. 
    \item We derive the expected squared ambiguity function of single-carrier FTN signals with respect to the given constellation alphabet. According to this result, the ranging performance is analyzed. Particularly, we define the accumulated ISI function for characterizing the impact of transmit ISI of FTN signaling in ranging, where its special oscillating response with respect to different delays is revealed. Furthermore, we show that this oscillating behavior is explicitly caused by the spectral aliasing, and verify that FTN signals have more robust ranging performance in terms of the expected squared ambiguity function. 
    \item We further evaluate the Doppler sensing performance of FTN signals and prove that the spectral aliasing will introduce undesired peaks in the expected squared ambiguity function along the Doppler dimension. As a result, no such undesired peaks will appear in the case of FTN transmissions with a sufficiently high symbol rate.
\end{itemize}

\emph{Notations:} The letters ${\mathbb{A}}$ and ${\mathbb{E}}$ denote the constellation set, the expectation operator, respectively; $\operatorname{Var}$ returns the variance of the underlying distribution; $(\cdot)^{\rm{T}}$ and $(\cdot)^{\rm{H}}$ denote the transpose and the Hermitian transpose for a matrix, respectively; 
$\det \left(\cdot\right)$ returns the determinant of a matrix;
``$ \Re \{\cdot\}$" returns the real part of a complex number; $\delta(\cdot)$ denotes the Dirac delta function; %$\lfloor x \rfloor$ is the floor function that returns the maximum integer number smaller than $x$;
$I \left({\bf a};{\bf b}\right)$ denotes the mutual information between $\bf a$ and $\bf b$;
${{{\bf{I}}_M}}$ denotes the identity matrix of size $M\times M$.
% ${\cal CN}\left(\mu,\sigma^2 \right)$ denotes the circularly symmetric Gaussian distribution with mean $\mu$ and variance $\sigma^2$. 

\section{System Models and Preliminaries}
In this study, we focus on the single-carrier FTN signaling using a real-valued shaping pulse $p\left(t\right)$ that is energy-normalized and band-limited, whose bandwidth{\footnote{The bandwidth in this paper is defined as the ``two-sided'' bandwidth, covering the frequency interval from negative frequency to positive frequency. The famous Nyquist no-ISI theorem considers the ``one-sided'' bandwidth, which is half of the ``two-sided'' bandwidth considered in the paper. Hence, it states that the symbol rate needs to be higher than twice of the bandwidth for avoiding ISI.}} is $W$. Let $\bf x$ be the transmitted symbol vector of length $N$, i.e., occupying $N$ time slots. Assume that the $n$-th entry of $\bf x$, i.e., $x_n$, for $1 \le n \le N$, equiprobably takes values from an energy-normalized complex constellation $\mathbb A$. Then, the transmitted signal is written as~\cite{Anderson2013FTN,rusek2009constrained}
\begin{align}
s\left(t\right)= \sqrt{E_s}\sum\limits_{n = 1}^{N} x_n p\left(t-n\xi T\right),
\label{FTN_s_t}
\end{align}
where $T$ is the Nyquist symbol period, $E_s$ is the average symbol energy, and $0 <\xi \le 1$ is the FTN compression factor. When $\xi=1$, the above transmitted signal degrades to the Nyquist signal. 

Let us consider the signal transmission over the time-invariant communication channel of the form
\begin{align}
h\left(t,\tau\right)= \sum\limits_{l = 1}^{L} h_l \delta\left(\tau-\tau_l\right),
\label{Channel}
\end{align}
which contains $L$ resolvable paths and each path has a resolvable delay $\tau_l$, i.e., $\tau_l \ne \tau_{l'}$, $\forall l,l' \in \left[1,L\right]$, and a fading coefficient $h_l$. 

According to~\eqref{Channel}, the received signal for communication can be written as
\begin{align}
r\left(t\right)= \sqrt{E_s}\sum\limits_{l = 1}^{L}\sum\limits_{n = 1}^{N} h_l x_n p\left(t-n\xi T-\tau_l\right)+n\left(t\right),
\label{FTN_r_t}
\end{align}
where $n\left(t\right)$ is the AWGN process with one-sided power spectral density (PSD) $N_0$.
After the matched-filtering with respect to $p\left(t\right)$, we obtain a set of sufficient statistics of the information symbol $\bf x$ given by
\begin{align}
{y_m} &= \int_{ - \infty }^\infty  {r\left( t \right){p^*}\left( {t - m\xi T} \right)} {\rm{d}}t\notag\\
&= \sum\limits_{l = 1}^L {{h_l}}\sqrt{E_s} \sum\limits_{n = 1}^N {{x_n}} g\left[ {n - m,{\tau _l}} \right]+\eta_m.
\label{FTN_y_m}
\end{align}
Here, $g\left[k,\tau\right]$ denotes the ISI between symbols that are apart in time by $kT+\tau$, which is given by
\begin{align}
g\left[ {k,\tau } \right] &\triangleq \int_{ - \infty }^\infty  {p\left( t \right){p^*}\left( {t + k\xi T + \tau } \right)} {\rm{d}}t\notag\\
&= \int_{ - \infty }^\infty  {{{\left| {{H_p}\left( f \right)} \right|}^2}\exp \left( {j2\pi f\left( {k\xi T + \tau } \right)} \right)} {\rm{d}}f.
\label{ISI}
\end{align}
where ${{H_p}\left( f \right)}$ is the Fourier transform of $p\left(t\right)$ and~\eqref{ISI} holds due to the Parseval's Theorem{\footnote{In this paper, we assume that the shaping pulse $p\left(t\right)$ is a Lebesgue integrable function such that its Fourier transform is well-defined.}}. In~\eqref{FTN_y_m}, $\eta_m$ is the corresponding \textit{colored} noise term satisfying ${\mathbb E} \left[\eta_m\eta_n^{*}\right]=g\left[n-m,0\right]N_0$.
Let $\bf y$ be the received symbol vector of length $N$, whose $m$-th element is given by $y_m$. According to~\eqref{FTN_y_m}, we have 
\begin{align}
{\bf y}=\sqrt{E_s}\sum\limits_{l = 1}^{L} h_l{\bf G}_l {\bf x}+{\bm \eta},
\label{x_y_io}
\end{align}
where ${\bm \eta} \triangleq \left[\eta_1,\eta_2,...,\eta_N \right]^{\rm T}$ is the colored noise sample vector and ${\bf G}_l$ is the effective channel matrix corresponding to the $l$-th resolvable path given by
\begin{align}
{{\bf{G}}_l} =\left[ {\begin{array}{*{20}{c}}
{g\left[ {0,{\tau _l}} \right]}&{g\left[ {1,{\tau _l}} \right]}& \cdots &{g\left[ {N - 1,{\tau _l}} \right]}\\
{g\left[ { - 1,{\tau _l}} \right]}&{g\left[ {0,{\tau _l}} \right]}&{}& \vdots \\
 \vdots &{}& \ddots & \vdots \\
{g\left[ {1 - N,{\tau _l}} \right]}& \cdots & \cdots &{g\left[ {0,{\tau _l}} \right]}
\end{array}} \right],
\label{G_l}
\end{align}
which is a real Toeplitz matrix. Furthermore, we have 
${\mathbb E} \left({\bm \eta}{\bm \eta}^{\rm H}\right)=N_0 {\bf{G}}_0$, where ${\bf{G}}_0$ is given by~\eqref{G_l} assuming $\tau_0=0$, which is a Hermitian Toeplitz matrix by definition. 
Notice that~\eqref{x_y_io} contains non-negligible ISI, and therefore an advanced equalizer is usually required for FTN signaling. Typical FTN equalizers include M-BCJR algorithms~\cite{Prlja2012MBCJR,Shuangyang2018Ungerboeck}, channel shortening based equalizers~\cite{Rusek2012CS,Li2020code}, and frequency-domain-equalization (FDE) based equalizers~\cite{Shinya2013FDE}. These equalizers have demonstrated good performance-complexity tradeoff in practical FTN transmissions~\cite{Shuangyang2020TDFD}.

For the sensing functionality, the ambiguity function of the transmitted signal $s\left(t\right)$ is defined as 
\begin{align}
{\cal AF}_s\left(\tau,\nu\right)\triangleq \int_{ - \infty }^\infty s\left(t\right)
s^{*}\left(t-\tau\right)e^{-j 2 \pi \nu \left(t-\tau\right)}{\rm d}t,
\label{AF_time}
\end{align}
where $\tau$ and $\nu$ are delay and Doppler offsets, respectively{\footnote{It should be noted that there is an alternative definition of the ambiguity function, i.e., ${\cal AF}_s\left(\tau,\nu\right)\triangleq \int_{ - \infty }^\infty s\left(t\right)
s^{*}\left(t-\tau\right)e^{-j 2 \pi \nu t}{\rm d}t$. The only difference is that~\eqref{AF_time} has an additional phase term. In fact, both definitions are widely considered and technically correct, where the additional phase term in~\eqref{AF_time} can be seen as the result that the signal is firstly shifted in frequency by $\nu$ and then shifted in time by $\tau$. }}. 
Alternatively, the ambiguity function can also be calculated from the frequency domain. Let $H_s (f)$ be the Fourier transform of $s\left(t\right)$. Then,~\eqref{AF_time} can be equivalently written by
\begin{align}
{\cal AF}_s\left( {\tau ,\nu } \right) = \int_{ - \infty }^\infty  {{H_s}\left( f \right)} {H_s^*}\left( {f - \nu } \right){e^{j2\pi f\tau }}{\rm{d}}f,
\label{AF_freq}
\end{align}
It is known that the squared ambiguity function essentially characterizes the  sensitivity of the transmitted signal at different delay and Doppler offsets when a matched-filtering based sensing receiver is applied{\footnote{To the best knowledge of authors, the sensing receiver design for FTN signaling is still a new topic. Therefore, we consider the conventional matched-filtering based sensing receiver in this paper.}}. 
In fact, the ambiguity function has a close relationship with the Cram{\'e}r–Rao bound that is the universal lower-bound for estimation problems. At low SNRs, the Cram{\'e}r–Rao bound primarily influenced by the sidelobes of the ambiguity function, whereas at high SNRs, it is governed by the curvature of the mainlobe~\cite{xiong2023snr}.
Therefore, in the framework of ISAC, we are interested in the average squared ambiguity function with respect to each pair of $\tau$ and $\nu$, i.e., ${\mathbb E}\left[ |{\cal AF}_s\left(\tau,\nu\right)|^2 \right]$, where the expectation is taken over the random vector of modulation symbols $\bf x$ after the square operation since ${\cal AF}_s\left(\tau,\nu\right)$ can have a complex value.
Further simplification on ${\mathbb E}\left[ |{\cal AF}_s\left(\tau,\nu\right)|^2 \right]$ is possible by considering~\eqref{FTN_s_t} and this will be studied in Section IV.  

\subsection{Assumptions and Preliminaries}
Before starting our analysis, it is important to first clearly introduce the assumptions and some frequently used preliminary knowledge in this subsection. Throughout this paper, we consider the following assumptions on the complex constellation $\mathbb A$. 

\textbf{Constellation Assumption} (\emph{Unit-Power Rotational Symmetric Constellation}):
Let $\mathbb A$ be a rotational symmetric constellation with unit-power. Then, we have 
\begin{align}
{\mathbb E}\left[a_i\right]=0, \quad {\mathbb E}\left[a_i^2\right]=0, \quad {\rm and} \quad {\mathbb E}\left[|a_i|^2\right]=1 ,
\label{Constellation_constraint}
\end{align}
where $a_i\in {\mathbb A}$ is a possible constellation point.

It should be noted that the commonly adopted constellations follow the above assumption generally, including the phase-shift keying (PSK), and QAM constellations, except for BPSK and $8$-QAM.   
A useful concept for our study is the \textit{kurtosis} of the constellation, which is defined as follows.

\textbf{Definition 1} (\emph{Constellation Kurtosis~\cite{liu2024ofdm,liu2024iceberg}}):
For a rotational symmetric constellation $\mathbb A$ with zero mean and unit power, its kurtosis is defined as
\begin{equation}
    \mu_4 \triangleq \frac{\mathbb{E}\left[|a_i-\mathbb{E}\left[a_i\right]|^4\right]}{\left\{\mathbb{E}\left[|a_i-\mathbb{E}\left[a_i\right]|^2\right]\right\}^2 } =  \mathbb{E}\left[|a_i|^4\right],\label{kurtosis}
\end{equation}
where $a_i\in {\mathbb A}$ is a possible constellation point.
\hfill $\lozenge$

As will be shown later, the kurtosis serves as an important metric for evaluating the sensing performance under random communication signals. 

We now introduce the concept of folded-spectrum that is widely used in the literature of FTN signaling. The definition of folded-spectrum is given as follows.

\textbf{Definition 2} (\emph{Folded-Spectrum~\cite{rusek2009constrained,kim2016properties}}):
Given the symbol rate $1/ \xi T$ and the underlying signaling pulse spectrum ${\left| {{H_p}\left( f \right)} \right|^2}$, the \emph{folded-spectrum} is a strictly band-limited frequency function defined as 
\begin{equation}
{\left| {{H_{{\rm{fo}}}}\left( f \right)} \right|^2} \buildrel \Delta \over = \sum\limits_{n =  - \infty }^\infty  {{{\left| {H_p\left( {f - \frac{n}{{\xi T}}} \right)} \right|}^2}} , \label{folded_spectrum}
\end{equation}
for $f \in \left[ { - \frac{1}{{2 \xi T}},\frac{1}{{2 \xi T}}} \right]$ and zero otherwise. \hfill $\lozenge$

Furthermore, it is convenient to define the twisted folded-spectrum as follows. 

\textbf{Definition 3} (\emph{Twisted Folded-Spectrum~\cite{Shuangyang2022FTNNOMA}}):
Given the symbol rate $1/ \xi T$ and the underlying signaling pulse spectrum ${\left| {{H_p}\left( f \right)} \right|^2}$, the \emph{twisted folded-spectrum} is a strictly band-limited frequency function defined as 
\begin{equation}
{\left| {{{ H}_{{\rm{tfo}}}}\left( f \right)} \right|^2} \triangleq {\left| {H_p\left( f \right)} \right|^2} - \sum\limits_{\scriptstyle n =  - \infty \hfill\atop
\scriptstyle n \ne 0\hfill}^\infty  {{{\left| {H_p\left( {f - \frac{n}{{\xi T}}} \right)} \right|}^2}} , \label{twisted_folded_spectrum}
\end{equation}
for $f \in \left[ { - \frac{1}{{2 \xi T}},\frac{1}{{2 \xi T}}} \right]$ and zero otherwise. \hfill $\lozenge$

\begin{figure*}[tp]
\label{Folded_spectrum_figs}
\centering
\subfigure[Folded-spectrum with $\xi_0=0.769$.]{
\begin{minipage}[t]{0.5\textwidth}
\centering
\includegraphics[scale=0.5]{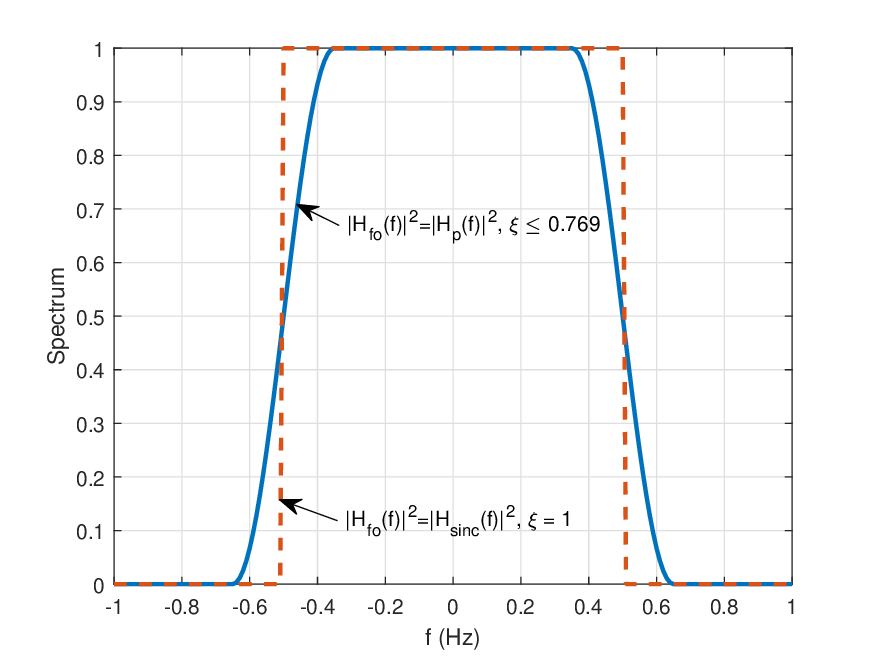}
\label{Folded_spectrum_fig}
\end{minipage}%
}%
\subfigure[Twisted folded-spectrum with $\xi_0=0.769$.]{
\begin{minipage}[t]{0.5\textwidth}
\centering
\includegraphics[scale=0.5]{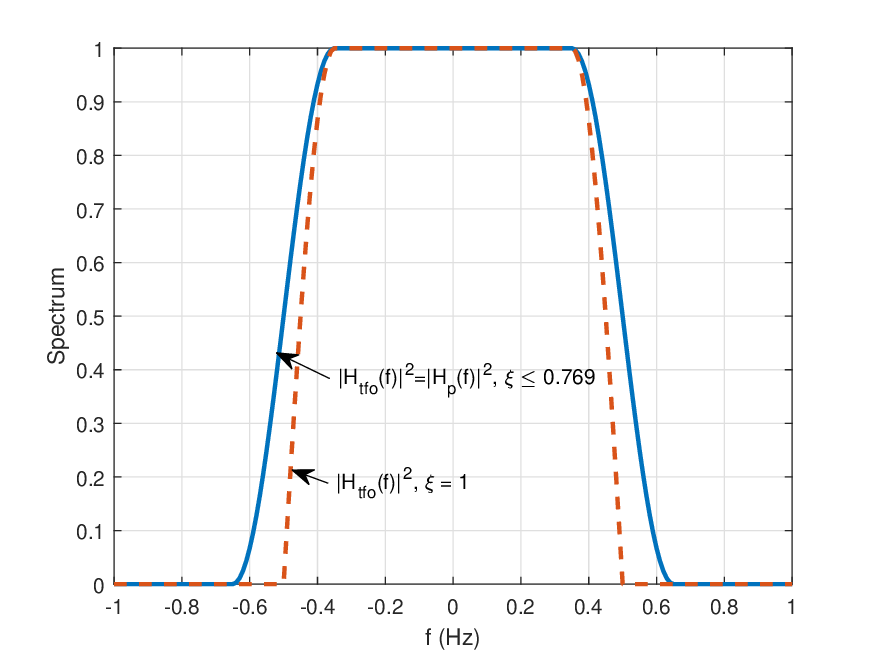}
%\caption{fig2}
\label{Twisted_Folded_spectrum_fig}
\end{minipage}%
}%
\caption{Illustrations of both folded-spectrum and twisted folded-spectrum with the saturation threshold $\xi_0=0.769$.}
\end{figure*}

The physical insight of the above definitions is important. In fact, both the folded-spectrum and the twisted folded-spectrum are two extreme cases considering the \textit{spectral aliasing}{\footnote{In fact, Nyquist signaling with sinc pulse can also avoid the spectral aliasing. However, since the sinc pulse is not realizable in practice, FTN signaling is therefore practically the only way for achieving no spectral aliasing~\cite{rusek2009constrained}.}} due to the mismatch of the symbol rate the and bandwidth of $p\left(t\right)$. Recall that  the bandwidth of 
$p\left(t\right)$ is $W$. The spectral aliasing happens when $W \ge \frac{1}{{\xi T}}$, and in this case the folded spectrum is the limiting case when the overlapped spectrum add up constructively. On the other hand, the twisted folded-spectrum indicates the case where the overlapped spectrum add up destructively. 
One interesting observation is that when $\xi$ is sufficiently small such that $W \le \frac{1}{{\xi T}}$, there will be no spectral aliasing and we have
${\left| {{{ H}_{{\rm{fo}}}}\left( f \right)} \right|^2}={\left| {{{ H}_{{\rm{tfo}}}}\left( f \right)} \right|^2}={\left| {{{ H}_p}\left( f \right)} \right|^2}$. We define the particular value of $\xi_0 \triangleq \frac{1}{{W T}}$ as the \textit{saturation threshold}, which indicates the maximum compression factor for the system to avoid the spectral aliasing.
More importantly, it was shown in~\cite{rusek2009constrained} that this saturation threshold is the maximum compression factor leading to the full exploitation of the system DoF. Note that for any $T$-orthogonal pulse $p\left(t\right)$, its bandwidth must satisfy that $W \ge \frac{1}{T}$, where the equality only achieves when $p\left(t\right)$ is a sinc pulse. Therefore, the effective DoF of FTN signaling is always larger than that of the Nyquist signaling under non-sinc shaping pulses.
We show the plots of the folded-spectrum and twisted
folded-spectrum in Fig.~\ref{Folded_spectrum_fig} and Fig.~\ref{Twisted_Folded_spectrum_fig}, where a RRC function with a roll-off factor $\beta=0.3$ is considered. Letting $T=1$, we have $W=1.3$ according to the definition of the RRC function, which yields the saturation threshold $\xi_0 \approx 0.769$.
As shown in figures, both the folded-spectrum and the twisted folded-spectrum align with the spectrum of the shaping pulse when $\xi\le \xi_0$. On the other hand, the folded-spectrum becomes the spectrum of the sinc pulse for $\xi=1$, while the twisted folded-spectrum for $\xi=1$ shows a degraded energy due to the destructive spectrum superposition.

In addition to the above discussions, it is also useful to discuss the so-called ``Dirichlet kernel'' of the form
\begin{align}
\sum\limits_{n = 1}^N {{e^{j2\pi xn\xi T}}} = {e^{j\pi x\left( {N + 1} \right)\xi T}}\frac{{\sin \left( {\pi xN\xi T} \right)}}{{\sin \left( {\pi x\xi T} \right)}}.
\label{Dirichlet_kernel}
\end{align}
The Dirichlet kernel is commonly studied in the field of Fourier analysis. Particularly, it can be shown that $\sum\nolimits_{n = 1}^N {{e^{j2\pi xn\xi T}}} $ is a periodic signal of $x$ with a period of $\frac{1}{\xi T}$. This property will be applied in the following discussions.

\subsection{Conditions for Fair Comparison}
The aim of this paper is to fairly evaluate the communication and sensing performance of FTN signals in comparison to the Nyquist counterpart. To this end, we consider the following conditions to ensure a fair comparison.
\begin{itemize}
    \item We consider both FTN and Nyquist signals have roughly the same PSD~\cite{rusek2009constrained,Shuangyang2022FTNNOMA}. This is ensured by setting $E_s=P \xi T$, where $P$ is the transmit power same for both FTN and Nyquist signals. In other words, the average symbol energy in FTN transmission is smaller than that of the Nyquist transmission, because more symbols are ``squeezed'' into the transmission duration for the FTN case, resulting in a more compact spectrum.
    \item We consider the normalized constrained capacity, i.e., maximum achievable spectral efficiency, \textit{without} water-filling for evaluating the communication performance, which inherently assumes Gaussian constellations and infinite codeword length. The meaning ``constrained capacity'' in our study refers to the constraints on the symbol rate, i.e., FTN rate or Nyquist rate~\cite{rusek2009constrained}. 
    \item We consider the expected normalized squared ambiguity function for evaluating the sensing performance. To make the problem mathematically trackable, a common approximation adopted in the literature is the following~\cite{liu2024ofdm,Hao2009design}
    \begin{align}
    {\mathbb E}\left[\frac{|{\cal AF}_s\left(\tau,\nu\right)|^2}{{|\cal AF}_s\left(0,0\right)|^2}\right] \approx \frac{{\mathbb E}\left[|{\cal AF}_s\left(\tau,\nu\right)|^2\right]}{{\mathbb E}\left[{|\cal AF}_s\left(0,0\right)|^2\right]}.
    \label{Approximation_normalization}
    \end{align}
    This simplifies the problem and allows us to focus on the analysis of ${\mathbb E}\left[|{\cal AF}_s\left(\tau,\nu\right)|^2\right]$. The motivation of this approximation is that ${|\cal AF}_s\left(0,0\right)|^2$ is roughly the sum symbol energy of the information symbols that stays invariant when $N$ is sufficiently large due to the law of large numbers.
    \item We assume that both FTN signals and Nyquist signals occupy roughly the same time and frequency resources for transmission. In other words, we require that FTN signal contains roughly $1/\xi$ more symbols compared to the Nyquist counterpart. This corresponds to the fact that both FTN and Nyquist signals are modulated by vectors of information symbols with different lengths but the same sum symbol energy. 
\end{itemize}
%%%%%%%%%%%%%%%%%%%
\section{Communication Performance Analysis}
To evaluate the communication performance, we study the normalized constrained capacity without water-filling in the following. Assume that the entries in $\bf x$ are i.i.d. circularly symmetric Gaussian variables with unit energy. The maximum achievable spectral efficiency of the considered FTN transmission is given by
\begin{align}
R\triangleq &\frac{1}{N \xi T W} I\left( {{\bf{y}};{\bf{x}}} \right) \notag\\
=&\frac{1}{N \xi T W}{\log _2}\det \left( {{{\bf{I}}_N} + \frac{{{E_s}}}{{{N_0}}}\sum\limits_{l = 1}^L \sum\limits_{l' = 1}^L{{h_lh_{l'}^{*}}{{\bf{G}}_l}{{\bf{G}}_{l'}^{\rm H}}{\bf{G}}_0^{ - 1}} } \right)\notag\\
&\quad\quad\quad\quad\quad\quad \quad\quad\quad\quad{\rm bits\;per\; second\; per \; Hz,}
\label{Mutual_info}
\end{align}
where the positive-definiteness of ${\bf G}_0$ in the asymptotic regime{\footnote{It was shown in~\cite{kim2016properties} that ${\bf G}_0$ is a full rank matrix for any strictly bandlimited or time-limited $p\left(t\right)$.}} 
is verified in~\cite{kim2016properties}.  
In fact, the underlying matrices involved in~\eqref{Mutual_info} have special properties that enables the efficient calculation of the spectral efficiency via Fourier analysis, which will be discussed in the following.

\subsection{Properties of ISI Matrices}
Let us first focus on ${\bf D}_{l,l'} \triangleq h_lh_{l'}^{*}{\bf G}_l {\bf G}_{l'}^{\rm H}$, $ \forall l,l' \in \left[1,L\right]$. In the asymptotic regime, i.e., $N \to \infty$, the $\left(n,m\right)$-th entry of ${\bf D}_{l,l'}$ is given by 
\begin{align}
D_{l,l'}\left[n,m\right]=h_lh_{l'}^{*}\sum\limits_{i = -\infty}^{\infty} g\left[i-n,\tau_l\right]g\left[i-m,\tau_{l'}\right].
\label{D_l_l_prime_entry}
\end{align}
From~\eqref{D_l_l_prime_entry}, one may easily verify that $D_{l,l'}\left[n,m\right]=D_{l,l'}\left[n+1,m+1\right]$, $\forall n,m$. Therefore, ${\bf D}_{l,l'}$ is a Toeplitz matrix in the asymptotic regime, whose Toeplitz coefficients are given by
\begin{align}
d_{l,l'}\left[n\right]=\sum\limits_{m =  - \infty }^\infty  h_lh_{l'}^{*} g\left[m, \tau_l\right]g\left[m-n, \tau_{l'}\right],
\label{d_ceof}
\end{align}
for $-\infty \le n \le \infty$. 
Furthermore, according to~\eqref{d_ceof}, it can be shown that $d_{l,l'}\left[n\right]  \ne d_{l,l'}^{*}\left[-n\right]$, for $l\ne l'$. Therefore, ${\bf D}_{l,l'}$ is not a Hermitian Toeplitz matrix for $l\ne l'$.

Even though ${\bf D}_{l,l'}$ may not be Hermitian in general, we can show that ${\bf T}_{l,l'}\triangleq{\bf D}_{l,l'} +{\bf D}_{l',l} $ is asymptotically Hermitian Toeplitz, for $l \ne l'$, by noticing that ${\bf D}_{l',l}={\bf D}_{l,l'}^{\rm H}$. More specifically, notice that the summation of Toeplitz matrices is also a Toeplitz matrix. Therefore, the Toeplitz coefficients of ${\bf T}_{l,l'}$ are given by
\begin{align}
t_{l,l'}\left[n\right]=&d_{l,l'}\left[n\right]+d_{l',l}\left[n\right], \notag\\
=&\sum\limits_{m =  - \infty }^\infty  h_lh_{l'}^{*} g\left[m, \tau_l\right]g\left[m-n, \tau_{l'}\right]\notag\\
&+\sum\limits_{m' =  - \infty }^\infty  h_{l'}h_l^{*} g\left[m', \tau_{l'}\right]g\left[m'-n, \tau_l\right],
\label{t_ceof}
\end{align}
for $-\infty \le n \le \infty$. 
Furthermore, it is not hard to verify that $t_{l,l'}\left[n\right]=t_{l,l'}^{*}\left[-n\right]$, and therefore ${\bf T}_{l,l'}$ is Hermitian. 

We have shown that both ${\bf D}_{l,l}$, and ${\bf T}_{l,l'}$ are Hermitian Toeplitz matrices in the asymptotic regime. Notice that 
\begin{align}
\sum\limits_{l = 1}^L \sum\limits_{l' = 1}^L {h_lh_{l'}^{*}}{{\bf{G}}_l}{{\bf{G}}_{l'}^{\rm H}}=\sum\limits_{l = 1}^L{\bf D}_{l,l}+\sum\limits_{i = 1}^{L} \sum\limits_{j = 1}^{i-1}{\bf T}_{i,j},
\end{align}
which is also a Hermitian Toeplitz matrix asymptotically. Note that ${\bf G}_0^{-1}$ is also a Hermitian Toeplitz matrix asymptotically~\cite{kim2016faster}. Hence, the matrix involved in the determinant in~\eqref{Mutual_info} is Hermitian Toeplitz asymptotically. As a result, the Szeg{\" o}'s Theorem can be applied for calculating~\eqref{Mutual_info}. This is studied in the following subsection.

\subsection{Spectral Efficiency Calculation based on Fourier Analysis}
To calculate the spectral efficiency, we first state the Szeg{\" o}'s Theorem as follows.

\textbf{Lemma 1} (\emph{The Szeg\"o's Theorem~\cite{gray2006toeplitz,simon2010szegHo}}):
Let ${\bf{V}}$ denote a size $N \times N$ Hermitian Toeplitz matrix ${\bf{V}}$, i.e.,
\begin{equation}
\setcounter{equation}{15}
\small
{\bf{V}} = \left[ {\begin{array}{*{20}{c}}
{{v_0}}&{{v_1}}& \cdots &{{v_{N - 1}}}\\
{{v_{ - 1}}}&{{v_0}}& \cdots &{{v_{N - 2}}}\\
 \vdots &{}& \ddots & \vdots \\
{{v_{1 - N}}}&{{v_{2 - N}}}& \cdots &{{v_0}}
\end{array}} \right],
\end{equation}
whose eigenvalues are given by $\left\{ {{\lambda _0},{\lambda _1}, \ldots {\lambda _{N - 1}}} \right\}$. Then, for an arbitrary continuous function $f_c(\cdot)$, we have
\begin{equation}
\mathop {\lim }\limits_{N \to \infty } \frac{1}{N}\sum\nolimits_{n = 0}^{N - 1} {{f_c}\left( {{\lambda _n}} \right) = \frac{1}{{2\pi }}\int_{ - \pi }^\pi  {{f_c}\left( {V\left( \omega  \right)} \right){\rm{d}}\omega } }  ,
\end{equation}
where ${V\left( \omega \right)}$ is the corresponding DTFT of the Toeplitz coefficients $\left\{ { \ldots ,{v_{ - 2}},{v_{ - 1}},{v_0},{v_1},{v_2}, \ldots } \right\}$ with $\omega \triangleq 2\pi f\xi T$, and it is given by
\begin{equation}
V\left( \omega  \right) = \sum\nolimits_{k =  - \infty }^\infty  {{v_k}{e^{ - jk\omega }}} .
\end{equation}
\hfill$\square$

To apply the Szeg{\" o}'s Theorem to~\eqref{Mutual_info}, we need to calculate the DTFTs of the Toeplitz coefficients of ${\bf G}_0$, ${\bf D}_{l,l}$, and ${\bf T}_{l,l'}$. 
According to~\cite{kim2016faster,Shuangyang2022FTNNOMA}, the DTFTs of the Toeplitz coefficients of ${\bf G}_0$ and ${\bf T}_l$ are calculated as
\begin{align}
G_0\left(2\pi f \xi T\right)=\sum\limits_{n =  - \infty }^\infty  g\left[n,\tau_0\right]\exp\left(-j 2 \pi n \xi Tf\right),
\label{g_DTFT}
\end{align}
\begin{align}
D_{l,l}\left(2\pi f \xi T\right)=&{|h_l|}^2\sum\limits_{n =  - \infty }^\infty  \sum\limits_{m =  - \infty }^\infty  g\left[m,\tau_l\right]g\left[m-n,\tau_l\right] \notag\\
&\exp\left(-j 2 \pi n \xi Tf\right),
\label{d_DTFT}
\end{align}
and 
\begin{align}
&T_{l,l'}\left(2\pi f \xi T\right)\notag\\
=&\sum\limits_{n =  - \infty }^\infty \bigg({h_lh_{l'}^{*}}\sum\limits_{m =  - \infty }^\infty  g\left[m,\tau_l\right]g\left[m-n,\tau_{l'}\right]\notag\\
&+{h_{l'}h_{l}^{*}}\sum\limits_{m' =  - \infty }^\infty g\left[m',\tau_{l'}\right]g\left[m'-n,\tau_{l}\right] \bigg) \exp\left(-j 2 \pi n\xi Tf\right),
\label{t_DTFT}
\end{align}
respectively. Note that ${\bf G}_0$, ${\bf D}_{l,l}$, and ${\bf T}_{l,l'}$ are Hermitian matrices. Therefore, $G_0\left(2\pi f \xi T\right)$, $D_{l,l}\left(2\pi f \xi T\right)$, and $T_{l,l'}\left(2\pi f \xi T\right)$ are of real values~\cite{gray2006toeplitz}. 
Further simplifications to~\eqref{g_DTFT},~\eqref{d_DTFT}, and~\eqref{t_DTFT} are possible by considering~\eqref{ISI} and the results are summarized in the following lemma.  

\textbf{Lemma 2} (\emph{DTFTs of Toeplitz Coefficients}):
In the case where ${{{\left| {{H_p}\left( {f } \right)} \right|}^2}}$ is symmetric, i.e. ${{{\left| {{H_p}\left( {f } \right)} \right|}^2}}={{{\left| {{H_p}\left( {-f } \right)} \right|}^2}}$, the DTFTs of Toeplitz coefficients for ${\bf G}_0$, ${\bf D}_{l,l}$, and ${\bf T}_{l,l'}$, are given by
\begin{align}
{G_0}\left( {2\pi f\xi T} \right) = \frac{1}{{\xi T}}\sum\limits_{n =  - \infty }^\infty  {{{\left| {{H_p}\left( {f - \frac{n}{{\xi T}}} \right)} \right|}^2}} ,
\label{g_DTFT_Freq}
\end{align}
\begin{align}
{D_{l,l}}\left( {2\pi f\xi T} \right) =& \frac{{|h_l|}^2}{{{{\left( {\xi T} \right)}^2}}}\sum\limits_{n =  - \infty }^\infty  {{{\left| {{H_p}\left( f-{\frac{n}{{\xi T}}} \right)} \right|}^2}{e^{j2\pi \frac{n}{{\xi T}}{\tau _l}}}} \notag\\
&\sum\limits_{m =  - \infty }^\infty  {{{\left| {{H_p}\left( {f - \frac{m}{{\xi T}}} \right)} \right|}^2}} {e^{ - j2\pi \frac{m}{{\xi T}}{\tau _l}}}
\label{d_DTFT_Freq},
\end{align}
and 
\begin{align}
&{T_{l,l'}}\left( {2\pi f\xi T} \right)\notag\\
=& 
\frac{2{\Re}\bigg\{h_l h_{l'}^{*}e^{j 2 \pi f (\tau_l-\tau_{l'})}\bigg\}}{{{\left( {\xi T} \right)}^2}}
\sum\limits_{n =  - \infty }^\infty  {{{\left| {{H_p}\left( f-{\frac{n}{{\xi T}}} \right)} \right|}^2}{e^{j2\pi \frac{n}{{\xi T}}{\tau _{l'}}}}} \notag\\
&\sum\limits_{m =  - \infty }^\infty  {{{\left| {{H_p}\left( {f - \frac{m}{{\xi T}}} \right)} \right|}^2}} {e^{ - j2\pi \frac{m}{{\xi T}}{\tau _{l}}}}
\label{t_DTFT_Freq},
\end{align}
respectively. 

\textbf{Proof}: The proofs are given in Appendix~A. \hfill$\square$

Notice from Lemma~2 that the channel delay has direct effect on ${D_{l,l'}}\left( {2\pi f\xi T} \right)$ and ${T_{l,l'}}\left( {2\pi f\xi T} \right)$, which essentially characterizes the eigenvalue distribution of ${\bf D}_{l,l}$ and ${\bf T}_{l,l'}$.
In fact, these DTFTs have a close connection with the folded-spectrum and twisted folded-spectrum defined in Section II-A, which is summarized 
in the following corollary.
%%%%%%%%%%
\begin{figure*}[tp]
\setcounter{equation}{28}
\begin{align}
R=\frac{1}{W}\int_{ - \frac{1}{{2\xi T}}}^{\frac{1}{{2\xi T}}} {{{\log }_2}} \left( {1 + \frac{P}{{{N_0}}}\frac{{\Upsilon \left( f \right)}}{{\sum\limits_{k =  - \infty }^\infty  {{{\left| {{H_p}\left( {f - \frac{k}{{\xi T}}} \right)} \right|}^2}} }}} \right){\rm{d}}f.
\label{Theorem1}
\end{align}
\begin{align}
\Upsilon \left( f \right) \triangleq& \sum\limits_{l = 1}^L {|{h_l}{|^2}} \sum\limits_{n =  - \infty }^\infty  {\sum\limits_{m =  - \infty }^\infty  {{{\left| {{H_p}\left( {f - \frac{n}{{\xi T}}} \right)} \right|}^2}} } {\left| {{H_p}\left( {f - \frac{m}{{\xi T}}} \right)} \right|^2}{e^{j2\pi \frac{{n - m}}{{\xi T}}{\tau _l}}}\notag\\
&+\!2\sum\limits_{l' = 1}^L {\sum\limits_{l'' = 1}^{l' - 1} {\Re \left\{ {{h_{l'}}h_{l''}^*e^{j 2 \pi f (\tau_{l'}-\tau_{l''})}}\right\}\sum\limits_{n' =  - \infty }^\infty  {{{\left| {{H_p}\left( {f \!-\! \frac{n'}{{\xi T}}} \right)} \right|}^2}{e^{j2\pi \frac{n'}{{\xi T}}{\tau _{l'}}}}} \!\!\sum\limits_{m' =  - \infty }^\infty  {{{\left| {{H_p}\left( {f \!-\! \frac{m'}{{\xi T}}} \right)} \right|}^2}} {e^{ - j2\pi \frac{m'}{{\xi T}}{\tau _{l''}}}} } } .
\label{Theorem1_add}
\end{align}
\setcounter{equation}{32}
\begin{align}
\Phi_{\rm UB}\left( f\right) \triangleq& \sum\nolimits_{l = 1}^L {|{h_l}{|^2}} |H_{\rm fo}\left(f\right)|^2+2\sum\nolimits_{l' = 1}^L {\sum\nolimits_{l'' = 1}^{l' - 1} {\Re \left\{ {{h_{l'}}h_{l''}^*e^{j 2 \pi f (\tau_{l'}-\tau_{l''})}}\right\} H_{\rm UB}\left(f\right)} } .
\label{Theorem1_add_UB}
\end{align}
\begin{align}
\Phi_{\rm LB} \left( f \right) \triangleq& \max\left\{\sum\nolimits_{l = 1}^L {|{h_l}{|^2}} |H_{\rm tfo}\left(f\right)|^2+2\sum\nolimits_{l' = 1}^L {\sum\nolimits_{l'' = 1}^{l' - 1} {\Re \left\{ {{h_{l'}}h_{l''}^*e^{j 2 \pi f (\tau_{l'}-\tau_{l''})}}\right\} H_{\rm LB}\left(f\right)} } ,0\right\}.
\label{Theorem1_add_LB}
\end{align}
\setcounter{equation}{23}
{\noindent} \rule[-10pt]{18cm}{0.05em}
\end{figure*}
%%%%%%%%%%

\textbf{Corollary 1} (\emph{DTFTs, Folded-Spectrum, and Twisted  Folded-Spectrum}):
The DTFT ${G_0}\left( {2\pi f\xi T} \right)$ given in~\eqref{g_DTFT_Freq} satisfies
\begin{align}
{G_0}\left( {2\pi f\xi T} \right) = \frac{1}{{\xi T}} {\left| {{{ H}_{{\rm{fo}}}}\left( f \right)} \right|^2},
\label{g_DTFT_Freq_FS}
\end{align}
for $f \in \left[ { - \frac{1}{{2 \xi T}},\frac{1}{{2 \xi T}}} \right]$.
Furthermore, the DTFT ${D_{l,l}}\left( {2\pi f\xi T} \right)$ given in~\eqref{d_DTFT_Freq} satisfies
\begin{align}
\frac{{{|h_l|}^2}}{{{{\left( {\xi T} \right)}^2}}}{\left| {{H_{{\rm{fo}}}}\left( f \right)} \right|^2}{\left| {{H_{{\rm{tfo}}}}\left( f \right)} \right|^2} \le {D_{l,l}}\left( {2\pi f\xi T} \right) \le \frac{{{|h_l|}^2}}{{{{\left( {\xi T} \right)}^2}}}{\left| {{H_{{\rm{fo}}}}\left( f \right)} \right|^4},
\label{D_DTFT_Freq_FS}
\end{align}
for $f \in \left[ { - \frac{1}{{2 \xi T}},\frac{1}{{2 \xi T}}} \right]$. Finally, the DTFT ${T_{l,l'}}\left( {2\pi f\xi T} \right)$ given in~\eqref{t_DTFT_Freq} satisfies
\begin{align}
&\frac{2{\Re\left\{h_l h_{l'}^* e^{j 2 \pi f\left(\tau_l-\tau_{l'}\right)}\right\}}}{{{{\left( {\xi T} \right)}^2}}}{\left| {{H_{{\rm{fo}}}}\left( f \right)} \right|^2}{{H_{{\rm{LB}}}}\left( f \right)}  \notag\\
\le &{T_{l,l'}}\left( {2\pi f\xi T} \right) 
\le \frac{2{\Re\left\{h_l h_{l'}^*e^{j 2 \pi f\left(\tau_l-\tau_{l'}\right)}\right\}}}{{{{\left( {\xi T} \right)}^2}}}{\left| {{H_{{\rm{fo}}}}\left( f \right)} \right|^2}{{H_{{\rm{UB}}}}\left( f \right)} ,
\label{T_DTFT_Freq_FS}
\end{align}
where 
\begin{align}
H_{\rm LB}\left(f\right)\triangleq \left\{
\begin{array}{cc}
    |H_{\rm tfo}\left(f\right)|^2, & {\rm if} \; {\Re\left\{h_l h_{l'}^*e^{j 2 \pi f\left(\tau_l-\tau_{l'}\right)}\right\}} \ge 0,  \\
    |H_{\rm fo}\left(f\right)|^2, & {\rm if} \; {\Re\left\{h_l h_{l'}^*e^{j 2 \pi f\left(\tau_l-\tau_{l'}\right)}\right\}} < 0,
\end{array}\right.
\label{H_LB}
\end{align}
and
\begin{align}
H_{\rm UB}\left(f\right)\triangleq \left\{
\begin{array}{cc}
    |H_{\rm fo}\left(f\right)|^2, & {\rm if} \; {\Re\left\{h_l h_{l'}^*e^{j 2 \pi f\left(\tau_l-\tau_{l'}\right)}\right\}} \ge 0,  \\
    |H_{\rm tfo}\left(f\right)|^2, & {\rm if} \; {\Re\left\{h_l h_{l'}^*e^{j 2 \pi f\left(\tau_l-\tau_{l'}\right)}\right\}} < 0,
\end{array}\right.
\label{H_UB}
\end{align}
respectively.

\textbf{Proof}: Notice that ${D_{l,l}}\left( {2\pi f\xi T} \right) = \frac{1}{{{{\left( {\xi T} \right)}^2}}}\sum\limits_{m =  - \infty }^\infty  {\sum\limits_{n =  - \infty }^\infty  {{{\left| {{H_p}\left( {f - \frac{n}{{\xi T}}} \right)} \right|}^2}{e^{j2\pi \frac{{n - m}}{{\xi T}}{\tau _l}}}} } {\left| {{H_p}\left( {f - \frac{m}{{\xi T}}} \right)} \right|^2}$. Therefore, the proof of Corollary 1 is straightforward by comparing~\eqref{folded_spectrum} and~\eqref{twisted_folded_spectrum} to~\eqref{g_DTFT_Freq},~\eqref{d_DTFT_Freq}, and~\eqref{t_DTFT_Freq}, and considering the Cauchy–Schwarz inequality~\cite{Shuangyang2022FTNNOMA}.  
\hfill$\blacksquare$

%%%%%%%%%%%%%%
With the above derivations, we are ready to calculate the spectral efficiency in~\eqref{Mutual_info} as shown in the following theorem. 

\textbf{Theorem 1} (\emph{Spectral Efficiency of FTN Signaling over Time-Invariant Multipath Channels}):
The spectral efficiency of FTN signaling over multipath channels under the system model in~\eqref{x_y_io} is given by~\eqref{Theorem1} as shown at the top of this page, where $P=E_s/\xi T$ is the average transmit power and $\Upsilon \left( {2\pi f\xi T} \right)$ is given in~\eqref{Theorem1_add}. Furthermore, it can be upper- and lower-bounded by
\setcounter{equation}{30}
\begin{align}
R\le \frac{1}{W}\int_{ - \frac{1}{{2\xi T}}}^{\frac{1}{{2\xi T}}} {{{\log }_2}\left( {1 + \frac{P}{{{N_0}}}\Phi_{\rm UB} \left(f\right)} \right)} {\rm{d}}f,
\label{Theorem1_lower}
\end{align}
and 
\begin{align}
R\ge \frac{1}{W}\int_{ - \frac{1}{{2\xi T}}}^{\frac{1}{{2\xi T}}} {{{\log }_2}\left( {1 + \frac{P}{{{N_0}}}\Phi_{\rm LB} \left(f\right)} \right)} {\rm{d}}f,
\label{Theorem1_upper}
\end{align}
respectively, where $\Phi_{\rm UB} \left(f\right) $ and $\Phi_{\rm LB} \left(f\right) $ are given in~\eqref{Theorem1_add_UB} and~\eqref{Theorem1_add_LB} at the top of this page, respectively. More importantly, the derived two bounds coincide with each other when the symbol rate is higher than the bandwidth of the shaping pulse, i.e., $\frac{1}{\xi T}\ge W$. In this case, we have 
\setcounter{equation}{34}
\begin{align}
R=\frac{1}{W}\int_{ - \frac{W}{{2}}}^{\frac{W}{{2}}} {{\log }_2}\left( 1 + \frac{P}{{{N_0}}} \left|\sum\limits_{l = 1}^L {{h_l e^{j2 \pi f \tau_l}}}\right|^2{{\left| {{H_p}\left( f \right)} \right|}^2}
\right) {\rm{d}}f.
\label{Theorem1_delay_independent}
\end{align}

\textbf{Proof}: The proofs are given in Appendix~B. \hfill$\square$

As shown in Theorem~1, the spectral efficiency of FTN signaling not only relates to the channel condition, but also depends on the connections between the spectrum of the shaping pulse and the symbol rate, i.e., the impact of spectral aliasing. Some interesting observations are summarized as follows.
\begin{itemize}
    \item We notice that the symbol rate potentially increase the DoF, i.e., the pre-log factor (integral range), of the system, while the channel delay affects only the effective SNR per frequency component for the transmission. Specifically, the effective SNR is determined by both the channel condition and the spectral aliasing with respect to the channel delay, where the impact of the latter is captured by the derived upper- and lower-bounds. Particularly, the upper- and lower-bounds correspond to the constructive and destructive superposition among spectrum components, which are achievable under particular channel delay and shaping pulse. In other words, the channel delay can introduce SNR variations for FTN signaling through spectral aliasing.

    \item There is a non-trivial tradeoff between the DoF and the potential SNR variation from spectral aliasing. Particularly, the potential SNR variation attains its maximum  when $\xi=1$, i.e., the Nyquist case{\footnote{In fact, our analysis also holds for slower-than-Nyquist signaling with $\xi >1$. However, such a signaling is rarely considered in practice and therefore we will not focus on the related discussion.}}, while the DoF is maximized when $\xi\le\frac{1}{WT}$. Similar tradeoff on SNR vs. DoF has also been reported in~\cite{Shuangyang2022FTNNOMA} under a different transmission scenario.

    \item In the limiting case, where the DoF is maximized, i.e., $\xi\le\frac{1}{WT}$, the potential SNR variation due to the spectral aliasing disappears. This observation aligns with the findings in~\cite{rusek2009constrained}. 
    \item Note that our above analysis is derived based on the properties of folded-spectrum and twisted folded-spectrum, which is not limited to only RRC pulses. In fact, our analysis can also extend to non-RRC pulses straightforwardly as long as both folded-spectrum and twisted folded-spectrum are well-defined. 

    \item It should be highlighted that our analysis is derived based on the time-invariant channel model, which may only hold for transmissions with a finite block length. In this context, the derived results can be viewed as a good approximation to the actual achievable rate given a sufficiently long block length, according to the asymptotic equivalence between the Toeplitz matrix and circulant matrix~\cite{Zhihui2017asymptotic,Benzin2019low}. Furthermore, this approximation becomes exact when a cyclic prefix (CP) longer than the channel delay spread is adopted in the system, such that the effective channel matrix becomes circulant.  
\end{itemize}

\subsection{Numerical Results of Communication Performance}
\begin{figure}[pt]
\centering
\includegraphics[width=0.4\textwidth]{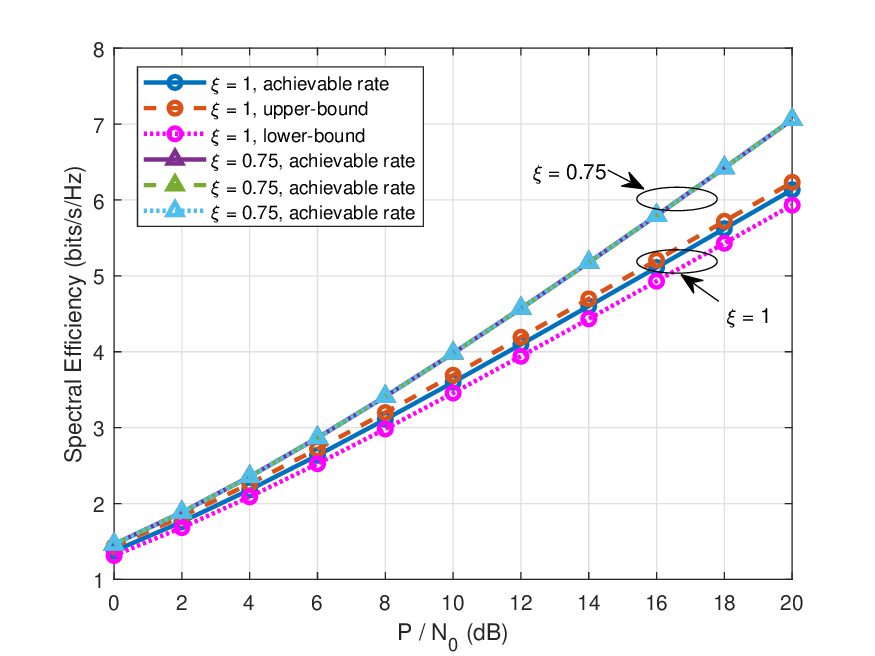}
	\caption{Numerical results of the spectral efficiency with $\xi=1$ and $\xi=0.75$, where the dashed and dot lines are the upper and lower bounds. Here, the RRC pulse with $\beta=0.3$ and $T=1$ is considered.}\label{SE_1}
\end{figure}
\begin{figure}[pt]
\centering
\includegraphics[width=0.4\textwidth]{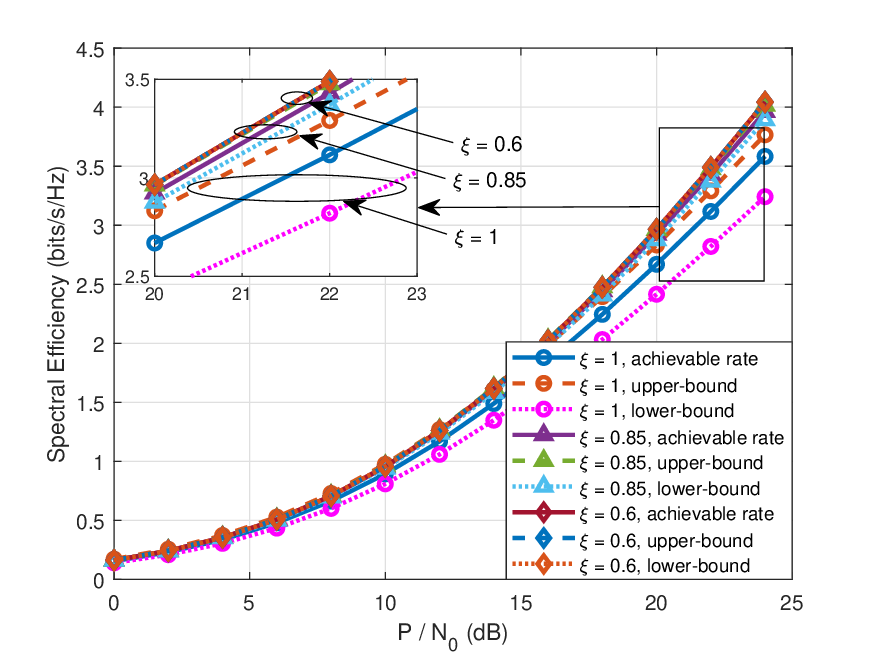}
	\caption{Numerical results of the ergodic spectral efficiency with $\xi=1$ and $\xi=0.85$, where the dashed and dot lines are the upper and lower bounds. Here, the RRC pulse with $\beta=0.3$ and $T=1$ is considered, and the channel has $L=3$ paths with a uniform power delay profile and the maximum delay is $\tau_{\max} = 2T$.}\label{SE_2}
\end{figure}

We present the numerical results of the derived spectral efficiency in this subsection. Under the assumption of i.i.d. circular symmetry Gaussian symbols, we define the SNR as $P/ N_0$ for evaluating the performance under the same PSD.
The spectral efficiency performance for both FTN and Nyquist signals is illustrated in Fig.~\ref{SE_1}, where a RRC shaping pulse with roll-off factor $\beta=0.3$ and $T=1$ is applied. In the simulation, a fixed channel with $L=3$ paths is considered, where the fading coefficients are given by $h_l=1/\sqrt{L}$, for $1 \le l \le L$, and the delays are given by $0$, $0.2T$, and $0.5T$, respectively. As observed, the actual spectral efficiency is accurately bounded by the derived upper- and lower-bounds for both FTN and Nyquist transmissions. Furthermore, we notice that the bounds coincide with each other for the FTN signaling with $\xi=0.75$, which is expected as $\xi_0 \approx 0.769$ for $\beta=0.3$. This indicates the system has the maximized DoF and no SNR gain from the spectral aliasing.

We illustrate the ergodic communication performance in Fig.~\ref{SE_2} with $\beta=0.3$ and $T=1$. Specifically, we consider the average spectral efficiency with $100000$ Monte Carlo trials, where the channel has $L=3$ paths, the fading coefficients are generated randomly following a uniform power delay profile, and the channel delays are uniformly distributed within $[0, \tau_{\max}]$ satisfying $\tau_{\max}=2T$. From the figure, we observe the same conclusions from the previous figure, where the SNR gain decreases with a higher symbol rate, while the DoF gain increases. This verifies the performance advantages of FTN signaling over Nyquist signaling in general time-invariant multipath fading channels. In addition, since the achievable rate of OFDM equals to that of single-carrier signaling under Gaussian constellations and optimal detection, it is expected that the considered single-carrier FTN signaling outperforms OFDM in terms of communication performance as well.
%%%%%%%%%%%%%%
\begin{figure*}[!hbp]
\rule{\textwidth}{0.5mm}
\setcounter{equation}{38}
\begin{align}
X(\tau)=& \int_{ - \infty }^\infty  {\int_{ - \infty }^\infty  {{{\left| {{H_p}\left( {{f_1}} \right)} \right|}^2}{{\left| {{H_p}\left( {{f_2}} \right)} \right|}^2}} } {e^{ - j2\pi \left( {{f_1} - {f_2}} \right)\tau }}A(f_1 - f_2, N, \xi){\rm{d}}{f_1}{\rm{d}}{f_2},
\label{Accumulated_ISI_final}
\end{align}
where
\begin{align}
%A(\Delta f, N, \xi)\triangleq\frac{{\sin \left( {\pi \Delta f\left( {N - 1} %\right)\xi T} \right)}}{{\sin \left( {\pi \Delta f\xi T} \right)}}\left( %{\frac{{\sin \left( {\pi \Delta fN\xi T} \right)}}{{\sin \left( {\pi \Delta f\xi %T} \right)}}\cos \left( {\pi \Delta f\xi T} \right) + \cos \left( {\pi \Delta %fN\xi T} \right)} \right).\\
A(\Delta f, N, \xi)\triangleq\frac{\sin^2(\pi N \Delta f \xi T)}{\sin^2 (\pi \Delta f \xi T)}.
\label{ISI_oscillation}
\end{align}
\centering
\end{figure*}
%%%%%%%%%%%%%%%%%
\section{Sensing Performance Analysis}
In this section, we study the sensing performance of FTN signaling by focusing on the ambiguity function. 
By substituting~\eqref{FTN_s_t} into~\eqref{AF_time}, we obtain
\begin{align}
\setcounter{equation}{29}
&{\cal AF}_s\left(\tau,\nu\right)\notag\\
=& {E_s}\sum\limits_{n = 1}^N {\sum\limits_{n' = 1}^N {{x_n}} } x_{n'}^*\int_{ - \infty }^\infty  p\left( {t - n\xi T} \right)\notag\\
&\quad\quad\quad\quad\quad\quad{p^*}\left( {t - n'\xi T - \tau } \right) {e^{ - j2\pi \nu \left( t-\tau \right) }}{\rm{d}}t\notag\\
=& {E_s}\sum\limits_{n = 1}^N {\sum\limits_{n' = 1}^N {{x_n}} } x_{n'}^* e^{ - j2\pi n' \nu \xi T }{\cal AF}_p\left( \left(n'-n\right)\xi T-\tau,\nu\right),
\label{AF_expanded}
\end{align}
where ${\cal AF}_p\left( \tau,\nu\right)$ is the ambiguity function of $p\left(t\right)$ with respect to the delay and Doppler offsets $\tau$ and $\nu$, respectively.
Furthermore, by taking the square norm of~\eqref{AF_expanded} and applying the expectation with respect to $\bf x$, we have
\begin{align}
&{\mathbb E}\left[\left|{\cal AF}_s\left(\tau, \nu\right)\right|^2\right]\notag\\
=& E_s^2\sum\limits_{n = 1}^N {\sum\limits_{n' = 1}^N {\sum\limits_{k = 1}^N {\sum\limits_{k' = 1}^N {{\mathbb E}\left[ {{x_n}x_{n'}^*x_k^*{x_{k'}}} \right]} } {e^{ - j2\pi n'\nu \xi T}}{e^{j2\pi k'\nu \xi T}}} } \notag\\
&{{\cal AF}_p}\left( {\left( {n' - n} \right)\xi T - \tau ,\nu } \right){\cal AF}_p^*\left( {\left( {k' - k} \right)\xi T - \tau ,\nu } \right).
\label{AF_squared}
\end{align}
According to~\eqref{Constellation_constraint}, it can be 
shown that
\begin{align}
{\mathbb E}\left[ {{x_n}x_{n'}^*x_k^*{x_{k'}}} \right]
= \left\{ \begin{array}{l}
{\mu _4},\quad n = n' = k = k',\\
1,\quad \; \; n = n',k = k',n \ne k,\\
1,\quad \;\; n = k,n' = k',n \ne n',\\
0,\quad \; \;{\rm else},
\end{array} \right. 
\label{kurtosis_expand}
\end{align}
where ${\mu _4}$ is the kurtosis defined in~\eqref{kurtosis}.
By substituting~\eqref{kurtosis_expand} into~\eqref{AF_squared}, we obtain
\begin{align}
&{\mathbb E}\left[\left|{\cal AF}_s\left(\tau, \nu\right)\right|^2\right]\notag\\
=& E_s^2N{\mu _4}{\left| {{\cal AF}_p\left( { - \tau ,\nu } \right)} \right|^2} \notag\\
&+ E_s^2 {\sum\limits_{n = 1}^N {\sum\limits_{\scriptstyle k = 1\hfill\atop
\scriptstyle  k \ne n\hfill}^N {{e^{ - j2\pi n\nu \xi T}}{e^{j2\pi k\nu \xi T}}} }  } {\left| {{\cal AF}_p\left( { - \tau ,\nu } \right)} \right|^2}\notag\\
&+E_s^2 \sum\limits_{m = 1}^N {\sum\limits_{\scriptstyle l = 1\hfill\atop
\scriptstyle l \ne m\hfill}^N {{{\left| {{\cal AF}_p\left( {\left( {l - m} \right)\xi T - \tau ,\nu } \right)} \right|}^2}} } \notag\\
=&E_s^2N\left({\mu _4}-2\right){\left| {{\cal AF}_p\left( { - \tau ,\nu } \right)} \right|^2}\notag\\
&+E_s^2 {\sum\limits_{n = 1}^N {e^{ - j2\pi n\nu \xi T}} {\sum\limits_{\scriptstyle k = 1}^N {{e^{j2\pi k\nu \xi T}}} }  } {\left| {{\cal AF}_p\left( { - \tau ,\nu } \right)} \right|^2}\notag\\
&+E_s^2 \sum\limits_{m = 1}^N {\sum\limits_{\scriptstyle l = 1}^N {{{\left| {{\cal AF}_p\left( {\left( {l - m} \right)\xi T - \tau ,\nu } \right)} \right|}^2}} } .
\label{AF_squared_der1}
\end{align}
It can be noticed from~\eqref{AF_squared_der1} that ${\mathbb E}\left[\left|{\cal AF}_s\left(\tau, \nu\right)\right|^2\right]$ is the squared mean plus the variance of the ambiguity function ${\cal AF}_s\left(\tau,\nu\right)$ with respect to the transmitted symbols, i.e.,
\begin{align}
{\mathbb E}\left[\left|{\cal AF}_s\left(\tau, \nu\right)\right|^2\right]=
\left|{\mathbb E}\left[{\cal AF}_s\left(\tau, \nu\right)\right]\right|^2+ \operatorname{Var}\left[{\cal AF}_s\left(\tau, \nu\right)\right],
\label{AF_squared_der2}
\end{align}
where 
\begin{align}
{\mathbb E}\left[\left|{\cal AF}_s\left(\tau, \nu\right)\right|^2\right]={E_s}\sum\limits_{n = 1}^N {{e^{ - j2\pi n\nu \xi T}}} {\cal AF}_p\left( { - \tau ,\nu } \right)
\label{AF_mean}
\end{align}
and 
\begin{align}
\operatorname{Var}\left[{\cal AF}_s\left(\tau, \nu\right)\right]=&E_s^2N\left({\mu _4}-2\right){\left| {{\cal AF}_p\left( { - \tau ,\nu } \right)} \right|^2}\notag\\
+&E_s^2 \sum\limits_{m = 1}^N {\sum\limits_{\scriptstyle l = 1}^N {{{\left| {{\cal AF}_p\left( {\left( {l - m} \right)\xi T - \tau ,\nu } \right)} \right|}^2}} },
\label{AF_var}
\end{align}
respectively. This is known as the ``Iceberg Theorem'' in~\cite{liu2024iceberg}, where the expected squared ambiguity function may be metaphorically explained as an iceberg in the sea, containing the iceberg part corresponding to the squared mean and the sea level part corresponding the variance.
%%%%%%%%%%%%%%%%%%%
\subsection{Performance Analysis of Ranging}
We first analyze the ranging performance of FTN signals based on the results in~\eqref{AF_squared_der1}. By letting $\nu=0$,~\eqref{AF_squared_der1} is reduced to  
\begin{align}
&{\mathbb E}\left[\left|{\cal AF}_s\left(\tau, 0\right)\right|^2\right]\notag\\
=&E_s^2{N^2}{\left| {{{\cal AF}_p}\left( { - \tau ,0} \right)} \right|^2}+ E_s^2 N\left( {{\mu _4} - 2} \right){\left| {{\cal AF}_p\left( { - \tau ,0} \right)} \right|^2}\notag\\
&+E_s^2\sum\limits_{m = 1}^N {\sum\limits_{l = 1}^N {{{\left| {{\cal AF}_p\left( {\left( {l - m} \right)\xi T - \tau ,0} \right)} \right|}^2}} } .
\label{AF_delay_only}
\end{align}
Clearly from~\eqref{AF_delay_only}, under given average symbol energy $E_s$, we notice that the symbol rate only affects the ranging performance through the accumulation of ISI in time. Let us define the \textit{accumulated ISI function} by 
\begin{align}
X\left(\tau\right) \triangleq \sum\limits_{m = 1}^N {\sum\limits_{l = 1}^N {{{\left| {{{\cal AF}_p}\left( {\left( {l - m} \right)\xi T - \tau ,0} \right)} \right|}^2}} } .
\label{Accumulated_ISI}
\end{align}
Further simplification on~\eqref{Accumulated_ISI} can be derived by expanding the  ambiguity function, which is summarized in the following lemma. 

\begin{comment}
\begin{figure}[pt]
\centering
\includegraphics[width=0.4\textwidth]{Fig/A_tau_N_xi.eps}
	\caption{$A(\Delta f, N, \xi)$ with $\xi=1$ and $\xi=0.8$, where $N=100$ and $T=1$.}\label{A_tau_N_xi}
\end{figure}
\end{comment}

\textbf{Lemma 3} (\emph{Accumulated ISI function}):
The accumulated ISI function defined in~\eqref{Accumulated_ISI} can be further simplified as~\eqref{Accumulated_ISI_final}, which is given at the bottom of this page.

\textbf{Proof}: The derivations are given in Appendix~C. \hfill$\square$

\begin{figure}[pt]
\centering
\includegraphics[width=0.4\textwidth]{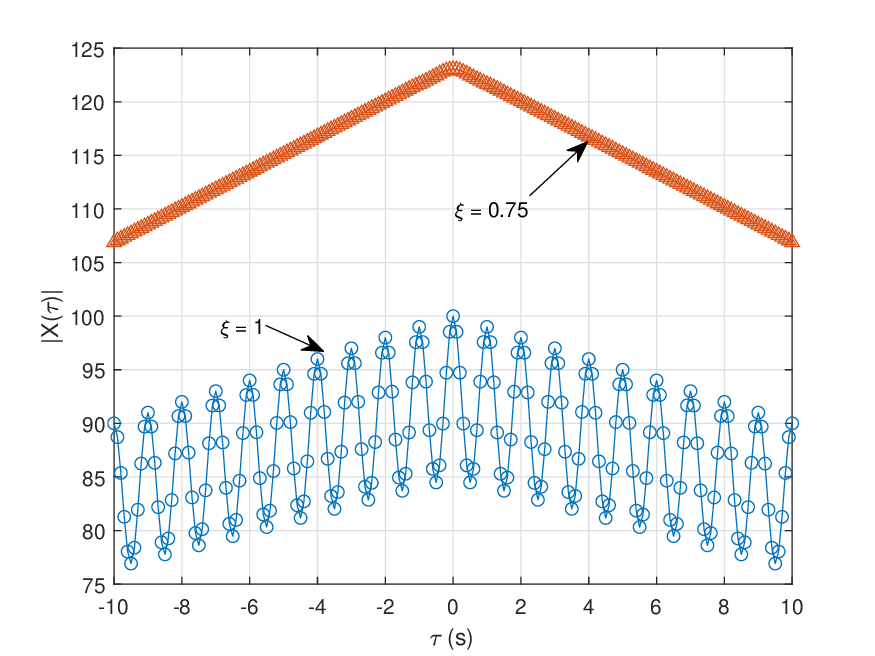}
	\caption{Comparison of the accumulated ISI function $X(\tau)$ with $\xi=1$ and $\xi=0.75$, where $N =100$, $T=1$, and a RRC shaping pulse with $\beta = 0.3$ is applied.}\label{x_tau_compare}
\end{figure}

From Lemma~3, we notice that the accumulated ISI function has a certain response due to the function of $A(\Delta f, N, \xi)$ given in~\eqref{ISI_oscillation}, which is a squared Dirichlet kernel, introducing the oscillation in its values. 
%We plot the $A(\Delta f, N, \xi)$ with $N=100$, $T=1$, and different $\xi$ in 
%Fig.~\ref{A_tau_N_xi}. 
Specifically, this function has peaky responses when $\Delta f$ is integer multiple of $\frac{1}{\xi T}$. 
%Particularly, we observe that a positive peaky response appears always at $\Delta f$ around zero no matter what $\xi$ is chosen, while other peaks in $A(\Delta f, N, \xi)$ may have either positive and negative values. 
Note that the frequency domain integration in~\eqref{Accumulated_ISI_final} involves the spectrum of the shaping pulse. This indicates that the the accumulated ISI function will have different behaviors depending on the bandwidth of the shaping pulse and the symbol rates, and the fundamental reason for this difference is precisely the effect of spectral aliasing.
Some key observations of the accumulated ISI function are as follows.
\begin{itemize}
    \item Under Nyquist signaling and shaping pulses with non-zero roll-off,  the spectral aliasing appears since $\xi =1 \ge \xi_0$. As a result, several peaks in $A(\Delta f, N, \xi)$ will be potentially included in the frequency domain integration in~\eqref{Accumulated_ISI_final}. Therefore, the accumulated ISI function $X(\tau)$ will suffer from fluctuations due to the oscillated values of $A(\Delta f, N, \xi)$.
    \item With a sufficiently high symbol rate, i.e., $\xi \le \xi_0$, there is no spectral aliasing, and therefore the frequency domain integration in~\eqref{Accumulated_ISI_final} can only include the positive peak at $\Delta f = 0$. As a result, the accumulated ISI function $X(\tau)$ experiences less fluctuation.
\end{itemize}
We plot the absolute value of $X(\tau)$ with different $\xi$ in Fig.~\ref{x_tau_compare}, where we consider $N =100$, $T=1$, and a RRC pulse with $\beta = 0.3$. In this case, $\xi_0 \approx 0.769$, and therefore we can see that $X(\tau)$ with $\xi = 0.75$ has a less fluctuated response and a generally higher absolute value compared to the Nyquist counterpart. It should be noted that this high absolute value does not directly lead to a higher sidelobe in the ambiguity function, after normalization in~\eqref{Approximation_normalization}.

Comparing the ranging performance with Nyquist signals, FTN signals generally lead to a less fluctuated expected squared ambiguity function along the delay axis, as evidenced by $X(\tau)$. This is potentially important for ranging with a large delay, e.g., remote sensing applications, because the corresponding sidelobe values are generally small. More detailed comparison of the ranging performance will be given in Section IV-C.
Furthermore, it should be mentioned that the ranging resolution can be improved by a larger signal bandwidth, which can be achieved by considering a larger roll-off factor $\beta$ or a smaller orthogonal period $T$. From a practical perspective, a smaller $T$ is more appealing for FTN signaling, because it does not require a very smaller $\xi$ to avoid the ISI, such that the detection complexity remains manageable.

%%%%%%%%%%%%%%%%%%%
\subsection{Performance Analysis of Doppler Sensing}
\begin{figure}[pt]
\centering
\includegraphics[width=0.4\textwidth]{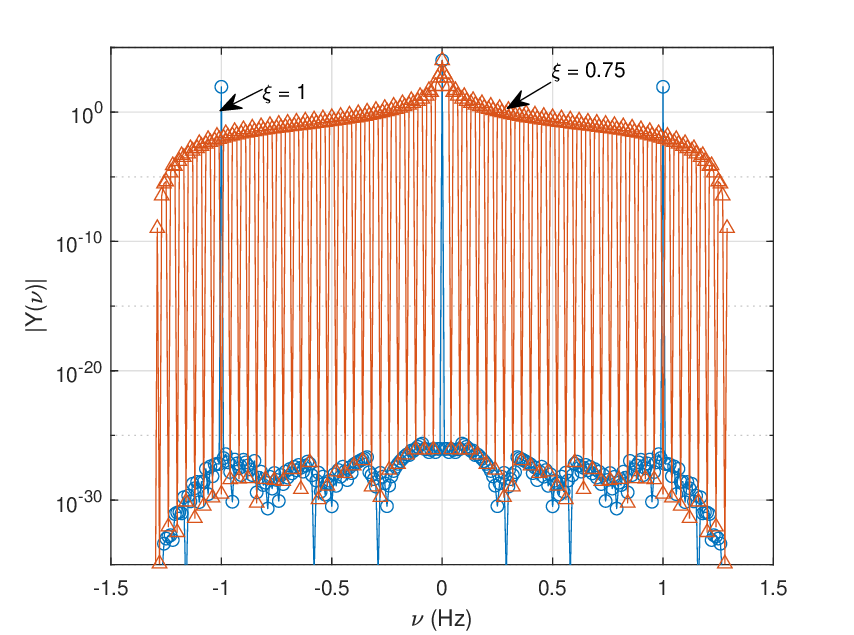}
	\caption{Comparison of the periodic Doppler variation function $Y(\nu)$ with $\xi=1$ and $\xi=0.75$, where $N =100$, $T=1$, and a RRC shaping pulse with $\beta = 0.3$ is applied.}\label{Y_nu_compare}
\end{figure}

We now turn our attention to Doppler sensing using FTN signals. By letting $\tau=0$,~\eqref{AF_squared_der1} is reduced to
\begin{align}
\setcounter{equation}{40}
&{\mathbb E}\left[\left|{\cal AF}_s\left(0, \nu\right)\right|^2\right]=\notag\\
&E_s^2N\left( {{\mu _4} \! - \! 2} \right){\left| {{\cal AF}_p\left( {0,\nu } \right)} \right|^2}\!+\!E_s^2\sum\limits_{m = 1}^N {\sum\limits_{l = 1}^N {{{\left| {{\cal AF}_p\left( {\left( {l \! -\! m} \right)\xi T,\nu } \right)} \right|}^2}} }\notag\\
&+E_s^2\sum\limits_{n = 1}^N {{e^{ - j2\pi n\nu \xi T}}} \sum\limits_{k = 1}^N {{e^{j2\pi k\nu \xi T}}} {\left| {{\cal AF}_p\left( {0,\nu } \right)} \right|^2}.
\label{AF_Doppler_only}
\end{align}
From~\eqref{AF_Doppler_only}, we define the \textit{Doppler-shifted accumulated ISI function} by
\begin{align}
{X'}\left(\nu\right) \triangleq \sum\limits_{m = 1}^N {\sum\limits_{l = 1}^N {{{\left| {{{\cal AF}_p}\left( {\left( {l - m} \right)\xi T ,\nu} \right)} \right|}^2}} } ,
\label{Doppler_Accumulated_ISI}
\end{align}
Similar to the derivation of Lemma~3,~\eqref{Doppler_Accumulated_ISI} can be further expanded as 
\begin{align}
X'(\nu)=&\int_{ - \infty }^\infty  \int_{ - \infty }^\infty  {{H_p^*}\left( {{f_1-\nu}} \right)}{{H_p}\left( {{f_2-\nu}} \right)}  \notag\\
&{{H_p}\left( {{f_1}} \right)}{{H_p^*}\left( {{f_2}} \right)}A(f_1 - f_2, N, \xi){\rm{d}}{f_1}{\rm{d}}{f_2},
\label{Doppler_Accumulated_ISI_expand}
\end{align}
where $A(\Delta f, N, \xi)$ is given in~\eqref{ISI_oscillation}. Similar to the accumulated ISI function in~\eqref{Accumulated_ISI}, the Doppler-shifted accumulated ISI function has fluctuating response with respect to different $\nu$, due to the oscillation from~\eqref{ISI_oscillation}, which can be mitigated with a smaller $\xi$.
Furthermore, we shall also define the \textit{periodic Doppler variation function}, which corresponds to $\left|{\mathbb E}\left[{\cal AF}_s\left(0, \nu\right)\right]\right|^2$, given by
\begin{align}
Y(\nu) \triangleq&\sum\nolimits_{n = 1}^N {{e^{ - j2\pi n\nu \xi T}}} \sum\nolimits_{k = 1}^N {{e^{j2\pi k\nu \xi T}}} {\left| {{\cal AF}_p\left( {0,\nu } \right)} \right|^2}\notag\\
=&\frac{{\sin^2 \left( {\pi \nu N\xi T} \right)}}{{\sin^2 \left( {\pi \nu \xi T} \right)}}{\left| {{\cal AF}_p\left( {0,\nu } \right)} \right|^2},
\label{P_Doppler_variation}
\end{align}
where~\eqref{P_Doppler_variation} holds due to~\eqref{Dirichlet_kernel}.
We observe from~\eqref{P_Doppler_variation} that the periodic Doppler variation function also
includes a squared Dirichlet kernel and consequently it potentially has several peaks at integer multiples of the period $\frac{1}{\xi T}$. We plot the periodic Doppler variation function with $N=100$, $T=1$ and different $\xi$ in Fig.~\ref{Y_nu_compare}, where a RRC pulse with $\beta=0.3$ is considered. 
We observe that in the Nyquist case, undesired peaks appear at integer times of the period $\frac{1}{\xi T}$, as expected. Notice again that $\xi_0 \approx 0.769$ for $\beta =0.3$. Therefore, we observe that no undesired peaks appear for $\xi = 0.75$, because there is no spectral aliasing.

From the above discussions, we are able to summarize the characteristics of Doppler sensing with different symbol rates as follows.
\begin{itemize}
    \item The spectral aliasing appears when Nyquist signaling is considered with shaping pulses having non-zero roll-off factors. In this case, the Doppler slice will not only have undesired peaks but also suffer from fluctuation due to the property of the Doppler-shifted accumulated ISI function.
    \item With a sufficiently high symbol rate, i.e., $\xi \le \xi_0$, there is no spectral aliasing, and therefore no undesired peaks appear in the ambiguity function. Furthermore, its value is more stable compared to the fluctuated response for the Nyquist counterpart.
\end{itemize}

%%%%%%%%%%%%%%%%

\subsection{Numerical Results of the Ambiguity Function}
We provide the numerical results in this subsection to verify the correctness of the derivation in the previous two subsections. Without loss of generality, we consider the QPSK constellation with unit kurtosis and set $T=1$. The shaping pulse is the RRC pulse with roll-off factor $\beta$. The length of the transmitted symbol vector for Nyquist transmission is $N=100$, and the length for the FTN transmission is adjusted according to the compression factor. In the results below, we will depict both  
${\mathbb E}\left[\frac{|{\cal AF}_s\left(\tau,\nu\right)|^2}{{|\cal AF}_s\left(0,0\right)|^2}\right] $ and $\frac{{\mathbb E}\left[|{\cal AF}_s\left(\tau,\nu\right)|^2\right]}{{\mathbb E}\left[{|\cal AF}_s\left(0,0\right)|^2\right]}$, which are denoted as ``actual'' and ``approx'', respectively. The value of ${\mathbb E}\left[\frac{|{\cal AF}_s\left(\tau,\nu\right)|^2}{{|\cal AF}_s\left(0,0\right)|^2}\right] $ is derived by using Monte Carlo simulations over $10000$ trials.

\begin{figure}[pt]
\centering
\includegraphics[width=0.4\textwidth]{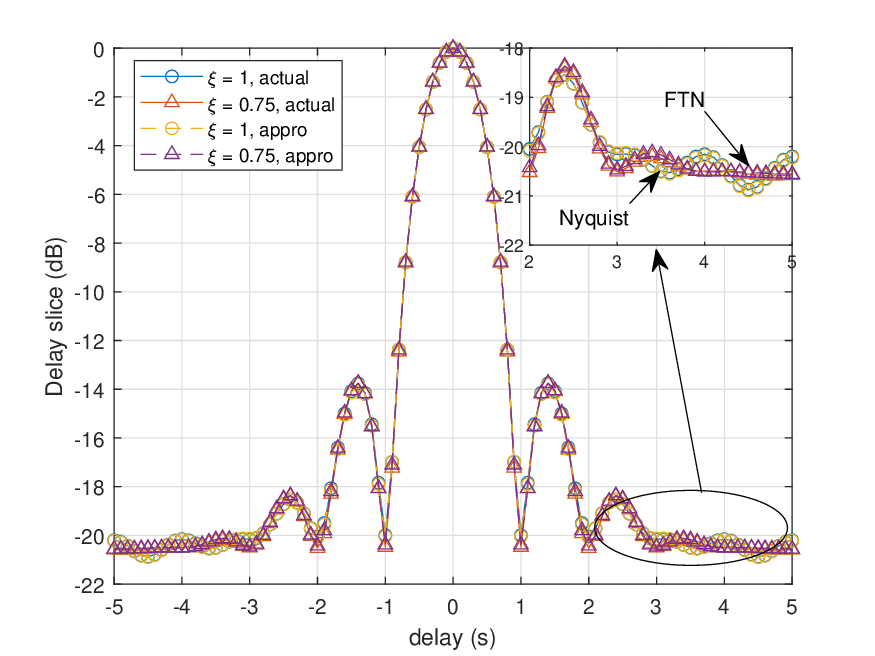}
	\caption{Comparison of the expected normalized squared ambiguity function along the delay dimension with $\xi=1$ and $\xi=0.75$, where $T=1$ and $\beta = 0.3$.}\label{AF_tau_compare}
\end{figure}

We plot the expected normalized squared ambiguity function along the delay dimension, i.e., the delay slice, in Fig.~\ref{AF_tau_compare}, where $\xi=1$ and $\xi=0.75$ are considered, and the FTN signal contains $100/0.75 \approx 133$ symbols. From the figure we observe that the considered approximation in~\eqref{Approximation_normalization} indeed yields a close match to the actual parameter of interest. Furthermore, we can see that both FTN and Nyquist signals share very similar behaviors overall, despite that the Nyquist signal yields more fluctuated response at 
larger delays. This is expected from our previous analysis, as the result of the spectral aliasing.

\begin{figure}[pt]
\centering
\includegraphics[width=0.4\textwidth]{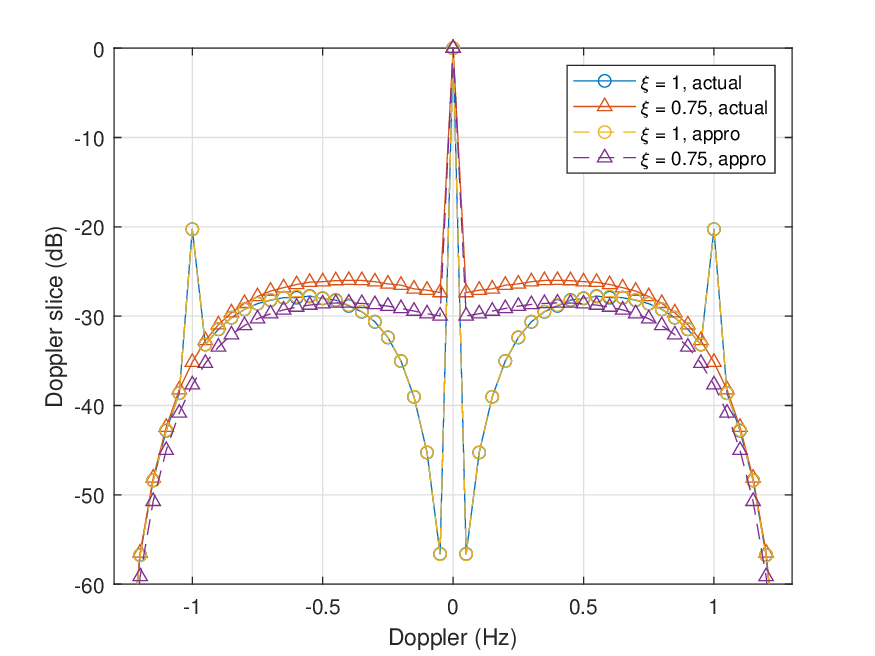}
	\caption{Comparison of the expected normalized squared ambiguity function along the Doppler dimension with $\xi=1$ and $\xi=0.75$, where $T=1$ and $\beta = 0.3$.}\label{AF_nu_compare}
\end{figure}

We plot the expected normalized squared ambiguity function along the Doppler dimension, i.e., the Doppler slice, in Fig.~\ref{AF_nu_compare}, where the same parameters as in the previous figure are considered. 
Similarly, we also observe a close match between the approximation and the actual value.
More importantly, we observe clearly undesired peaks appeared for the Nyquist transmissions, while no such peaks occurred in the FTN case with $\xi = 0.75$. 
This observation verifies the advantages of FTN signals for sensing comparing to the Nyquist counterpart.

\begin{figure}[pt]
\centering
\includegraphics[width=0.4\textwidth]{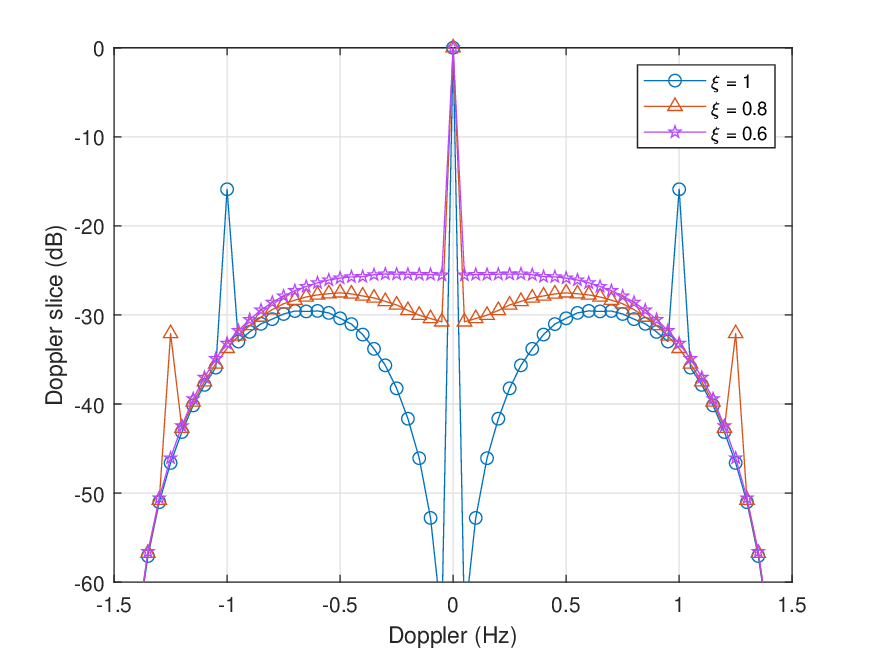}
	\caption{Comparison of the expected normalized squared ambiguity function along the Doppler dimension with $T=1$ and $\beta = 0.5$, where $\xi=1$, $\xi=0.8$, and $\xi=0.6$ are considered.}\label{AF_nu_xi_compare}
\end{figure}

We evaluate the Doppler slice with different compression factors in Fig~\ref{AF_nu_xi_compare}, where $T=1$ and $\beta = 0.5$ are considered. 
From the figure, we observe that cases with $\xi= 1$ and $\xi= 0.8$ suffer from the undesired peaks that appear at $\frac{1}{\xi T}$.
On the other hand, no undesired peaks appear for the $\xi= 0.6$ since $\xi <\xi_0 =\frac{2}{3}$ in the considered case. These observations imply that the strength of the undesired peaks reduces with a higher symbol rate until the saturation threshold $\xi_0$ is achieved. 

\subsection{Numerical Results of the Doppler Sensing Performance}
\begin{figure}[pt]
\centering
\includegraphics[width=0.4\textwidth]{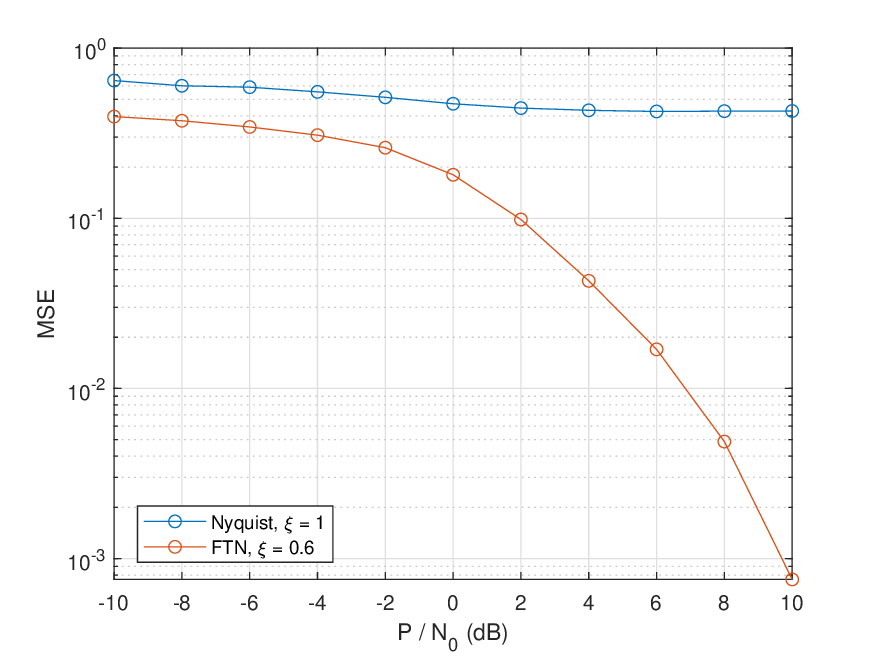}
	\caption{Comparison of Doppler sensing performance with $T=1$ and $\beta = 0.5$, where both Nyquist signal and FTN signal with $\xi=0.6$ are considered, and the QPSK constellation is applied. Here, one strong target and one weak target ($15\%$ of the reflectivity strength) are considered with normalized Doppler at $0.5$ Hz and $-0.4$ Hz.  }\label{Actual_Sensing_performance}
\end{figure}
Due to the space limitation, we only present the Doppler sensing performance in this subsection. We consider the Doppler sensing of two targets with strong and weak reflectivity strengths, where the strong target has normalized Doppler $0.5$ Hz while the weak target has normalized Doppler $-0.4$ Hz. We assume that the weak target has $15\%$ of the reflectivity strength compared to the strong target, such that its response may be masked by the sidelobe of the strong target after matched filtering-based estimation. 
We consider the Doppler sensing performance of both Nyquist signal and FTN signal with $\xi=0.6$, where $T=1$, $\beta = 0.5$, and QPSK constellation are applied. From the figure, we observe that the Nyquist signal fails to provide a reliable sensing performance in terms of the mean squared error (MSE), while FTN signal shows a good sensing performance that improves with an increased SNR. This observation aligns with our expectation and verifies the advantages of FTN signaling for sensing.
In addition, it was shown in~\cite{liu2024ofdm} that OFDM exhibits worse Doppler sensing performance than single-carrier signaling. Therefore, 
it is expected that the considered FTN signaling will outperform OFDM for Doppler sensing as well.
However, it should be highlighted that the simulation considers the adoption of the RRC pulse, which is not optimized for improving the sensing performance. With well-optimized pulses, it may be possible that the undesired peak in the expected normalized squared ambiguity function of Nyquist signaling is reduced, such that the Doppler sensing performance of Nyquist signaling can be improved.

\section{Potential Extensions and Future Works }
We have shown the advantages of single-carrier FTN signaling for both communications and sensing. In this section, we aim to provide high-level discussions on the extension of these results to other transmission schemes. 

\subsection{Extension to Multi-carrier FTN Signaling}
It should be noted that the term ``multi-carrier FTN signaling'' can be misleading in the sense that effectively all signaling schemes are implemented in the time domain in practice. Therefore, we highlight that we only consider the multi-carrier FTN signaling based on OFDM with reduced spacing between adjacent OFDM symbols and subcarriers~\cite{Dasalukunte2011multicarrier} in this paper. The ISAC performance analysis for multi-carrier FTN signaling is an interesting extension of the previous discussions. Note that the multi-carrier FTN signaling is a two-dimensional (2D) waveform and there is no signal that has strictly limited durations in both time and frequency. Therefore, it is important to adopt a reasonable method to measure time and bandwidth occupied for the transmission, such that the potential aliasing in both time and frequency can be well-defined. Then, one shall apply the derivations in this paper for evaluating the ISAC performance by considering the impact of the aliasing. Although we cannot provide the details on this discussion in this paper, it is expected that the total DoF of the multi-carrier FTN signaling is determined by the product of the time and bandwidth occupied according to the Shannon's $2WT$ theorem. Furthermore, its expected squared ambiguity function may have undesired peaks due to the potential aliasing in both time and frequency.

\subsection{Extension to MIMO Transmissions}
MIMO transmission serves as an important enabling technology for high-throughput transmissions in future wireless networks. The application of FTN signaling in MIMO transmissions can therefore further improve the achievable rate of MIMO transmissions. While the achievable rates of FTN-MIMO over AWGN channels have been studied in~\cite{Rusek2007MIMOFTN, Zichao2023FTN}, its performance over fading channels is still under-explored. The analysis of communication performance in this paper can be potentially extended to the MIMO case with ``on-grid'' angles-of-departure (AoDs) and angles-of-arrival (AoAs), where the interference from different antenna pairs can be effectively eliminated by beamforming. In this case, the FTN-MIMO transmission may be equivariant to the system model in~\eqref{x_y_io} with different distributions on the fading coefficients. Therefore, the method adopted in this paper may also be applicable.
On the other hand, the sensing performance for FTN-MIMO may be studied by deriving the three-dimensional (3D) ambiguity function with respect to the delay, Doppler, and angle. Since typical FTN transmission only increases the symbol rate in the time and frequency domain, the expected squared ambiguity function may exhibit similar performance in delay and Doppler dimensions. However, the angular sensing performance for FTN-MIMO transmissions still requires further study.

\subsection{Extension to Multi-User Transmissions}
Most studies of FTN signaling in multiple access transmissions assume only one resolvable path~\cite{Shuangyang2022FTNNOMA,Zichao2023capacity}, and the impact of multipath transmission remains under-explored.
The communication analysis of this paper can be extended to the uplink multiple access transmission by considering the application of successive interference cancellation (SIC) detection. Specifically, the SIC detection treats the interference from undecoded users as noise and refines the channel observations by excluding the contribution from the decoded users. Therefore, the communication achievable rate can be analyzed is a layer-by-layer manner similar to~\cite{Shuangyang2022FTNNOMA}, where the effective input-output relation for each layer is equivalent to the system model in~\eqref{x_y_io} that can be analyzed following the derivations in this paper.

\section{Conclusions}
In this paper, the application of single-carrier FTN signaling for ISAC is studied. According to our analysis, the following key advantages are identified: 1) single-carrier FTN signaling can effectively avoid the spectral aliasing and increase the DoF of the signal transmission; 2) single-carrier FTN signaling can mitigate the fluctuation of the ambiguity function sidelobe in the delay dimension; 3) single-carrier FTN signaling can avoid the undesired peaks of the ambiguity function in the Doppler dimension.
We conclude that these findings come from the fact that FTN signaling can effectively avoid the spectral aliasing due to the mismatch between the symbol rate and shaping pulse's bandwidth.
In addition to these findings, we also acknowledge the practical drawbacks on FTN signaling for ISAC, including the high implementation complexity and lack of good channel codes. These important practical issues require further studies. 
%%%%%%%%%%%%%%%%
\appendices

\section{Proof of Lemma~2}
By substituting~\eqref{ISI} into~\eqref{g_DTFT}, we have~\cite{Shuangyang2022FTNNOMA}
\begin{align}
{G_0}\left( {2\pi f\xi T} \right) &= \int_{ - \infty }^\infty  {{{\left| {{H_p}\left( \lambda  \right)} \right|}^2}{e^{j2\pi \lambda {\tau _0}}}\sum\limits_{n =  - \infty }^\infty  {{e^{j2\pi n\xi T\left( {\lambda  - f} \right)}}} } {\rm{d}}\lambda \notag\\
&= \frac{1}{{\xi T}}\sum\limits_{n =  - \infty }^\infty  {{{\left| {{H_p}\left( {f - \frac{n}{{\xi T}}} \right)} \right|}^2}} 
\label{G_l_Freq},
\end{align}
where~\eqref{G_l_Freq} holds due to the Poisson summation formula.
Similar to the above derivation, by considering~\eqref{ISI}, we obtain
\begin{align}
&{D_{l,l'}}\left( {2\pi f\xi T} \right) \notag\\
= & h_l h_l^{*}\sum\nolimits_{m =  - \infty }^\infty  g \left[ {m,{\tau _l}} \right]\sum\nolimits_{n =  - \infty }^\infty  {{e^{-j2\pi n\xi Tf}}} \notag\\
&\int_{ - \infty }^\infty  {{{\left| {{H_p}\left( \lambda  \right)} \right|}^2}{e^{j2\pi \lambda \left( {\left( {m - n} \right)\xi T + {\tau _{l'}}} \right)}}} {\rm{d}}\lambda \notag\\
=& h_l h_l^{*} \sum\nolimits_{m =  - \infty }^\infty  g \left[ {m,{\tau _l}} \right]\notag\\
&\int_{ - \infty }^\infty  {{{\left| {{H_p}\left( \lambda  \right)} \right|}^2}{e^{j2\pi \lambda \left( {m\xi T + {\tau _{l'}}} \right)}}\sum\nolimits_{n =  - \infty }^\infty  {{e^{ - j2\pi \left( {\lambda  + f} \right)n\xi T}}} } {\rm{d}}\lambda \notag\\
=&\frac{h_l h_l^{*}} {{\xi T}}\sum\nolimits_{n =  - \infty }^\infty  {{{\left| {{H_p}\left( {\frac{n}{{\xi T}} - f} \right)} \right|}^2}{e^{ - j2\pi \left( {f - \frac{n}{{\xi T}}} \right){\tau _{l'}}}}} \notag\\
&\sum\nolimits_{m =  - \infty }^\infty  {\int_{ - \infty }^\infty  {{{\left| {{H_p}\left( \lambda  \right)} \right|}^2}{e^{j2\pi \left( {\lambda  - f} \right)m\xi T}}{e^{j2\pi \lambda {\tau _l}}}{\rm{d}}\lambda } }  \notag\\
=&\frac{h_l h_l^{*}}{{{{\left( {\xi T} \right)}^2}}}\sum\nolimits_{n =  - \infty }^\infty  {{{\left| {{H_p}\left( {\frac{n}{{\xi T}} - f} \right)} \right|}^2}{e^{j2\pi \frac{n}{{\xi T}}{\tau _{l'}}}}}  \notag\\
&\sum\nolimits_{m =  - \infty }^\infty  {{{\left| {{H_p}\left( {f - \frac{m}{{\xi T}}} \right)} \right|}^2}} {e^{ - j2\pi \frac{m}{{\xi T}}{\tau _l}}}
e^{j 2 \pi f \left(\tau_{l}-\tau_{l'}\right)}
\label{D_l_Freq}.
\end{align}
Thus, in the case where ${\left| {{H_p}\left( f \right)} \right|^2}$ is symmetric,~\eqref{d_DTFT_Freq} can be obtained by letting $l=l'$. 
Finally, by noticing that 
\begin{align}
{T_{l,{l'}}}\left( {2\pi f\xi T} \right)=&{D_{l,l'}}\left( {2\pi f\xi T} \right)+{D_{l',{l}}}\left( {2\pi f\xi T} \right)\notag\\
=&2 {\Re}\left\{{D_{l,l'}}\left( {2\pi f\xi T} \right)\right\}
\label{T_l_Freq},
\end{align}
we can obtain ${T_{l,{l'}}}\left( {2\pi f\xi T} \right)$ as shown in~\eqref{t_DTFT_Freq} by similar derivations given in~\eqref{D_l_Freq}.
\hfill $\blacksquare$

%%%%%%%%%%
\section{Proof of Theorem~1}
According to Lemma~1, we have
\begin{align}
I\left( {{\bf{y}};{\bf{x}}} \right)
=& \frac{1}{{2\pi }}\int_{ - \pi }^\pi  {{\log }_2}\bigg(1 +\frac{{{E_s}}}{{{N_0}}}\big(\sum\limits_{l = 1}^L {D_{l,l}}\left( \omega  \right)+\notag\\
&\sum\limits_{l' = 1}^L \sum\limits_{l'' = 1}^{l'-1} {T_{l',l''}}\left( \omega  \right)\big)G_0^{ - 1}\left( \omega  \right)  \bigg) {\rm{d}}\omega \notag\\
= &{\xi T}\int_{ - \frac{1}{{2\xi T}}}^{\frac{1}{{2\xi T}}}  {{\log }_2}\bigg(1 +\frac{{{E_s}}}{{{N_0}}}\big(\sum\limits_{l = 1}^L {D_{l,l}}\left(2\pi f \xi T  \right)+\notag\\
&\sum\limits_{l' = 1}^L \sum\limits_{l'' = 1}^{l'-1}\! {T_{l',l''}}\!\left( 2\pi f \xi T \right)\!\big)G_0^{ - 1}\left( 2\pi f \xi T \right)  \bigg) \!{\rm{d}}f.
\label{Theorem1_der1}
\end{align}
Then, by substituting~\eqref{g_DTFT_Freq},~\eqref{d_DTFT_Freq}, and~\eqref{t_DTFT_Freq} into~\eqref{Theorem1_der1}, and considering the normalization, we arrive at~\eqref{Theorem1}.  
Furthermore, by substituting~\eqref{T_DTFT_Freq_FS} into~\eqref{Theorem1}, we can obtain~\eqref{Theorem1_lower} and~\eqref{Theorem1_upper}. Finally,~\eqref{Theorem1_delay_independent} holds according to the connections among signal spectrum, folded-spectrum, and the twisted folded-spectrum and the fact that $ \left|\sum\limits_{l = 1}^L {{h_l e^{j2 \pi f \tau_l}}}\right|^2=\sum\limits_{l = 1}^L {|{h_l}{|^2}} +2\sum\limits_{l' = 1}^L {\sum\limits_{l'' = 1}^{l' - 1} {\Re \left\{ {{h_{l'}}h_{l''}^*e^{j 2 \pi f (\tau_{l'}-\tau_{l''})}}\right\} } }$.
\hfill $\blacksquare$

%%%%%%%%%%%%%%%%
\section{Proof of Lemma~3}
Note that~\eqref{Accumulated_ISI} can be rewritten as
\begin{align}
X(\tau)&=\sum\nolimits_{m = 1 - N}^{N - 1} {\left( {N - \left| m \right|} \right){{\left| {{{\cal AF}_p}\left( {m\xi T - \tau ,0} \right)} \right|}^2}} \notag\\
=&\int_{ - \infty }^\infty  {\int_{ - \infty }^\infty  {{{\left| {{H_p}\left( {{f_1}} \right)} \right|}^2}{{\left| {{H_p}\left( {{f_2}} \right)} \right|}^2}{e^{ - j2\pi \left( {{f_1} - {f_2}} \right)\tau }}} } \notag\\
&\sum\nolimits_{m = 1 - N}^{N - 1} {\left( {N - \left| m \right|} \right)} {e^{j2\pi \left( {{f_1} - {f_2}} \right)m\xi T}}{\rm{d}}{f_1}{\rm{d}}{f_2}.
\label{Deriving_ISI1}
\end{align}
Notice that 
\begin{align}
&\sum\limits_{m = 1 - N}^{N - 1} {\left( {N - \left| m \right|} \right)} {e^{j2\pi \left( {{f_1} - {f_2}} \right)m\xi T}}\notag\\
=&N + \sum\limits_{m = 1}^{N - 1} {\left( {N - m} \right)} \left( {{e^{j2\pi \left( {{f_1} - {f_2}} \right)m\xi T}} + {e^{ - j2\pi \left( {{f_1} - {f_2}} \right)m\xi T}}} \right)\notag\\
=&N + 2\sum\limits_{m = 1}^{N - 1} {\left( {N - m} \right)} \cos \left( {2\pi \left( {{f_1} - {f_2}} \right)m\xi T} \right).
\label{Deriving_ISI2}
\end{align}
Furthermore, for a given $\Delta f$, we have 
\begin{align}
&\sum\nolimits_{m = 1}^{N - 1} {\left( {N - m} \right)} \cos \left( {2\pi \Delta f m\xi T} \right)\notag\\
=&\sum\nolimits_{m = 1}^{N - 1} {\sum\nolimits_{l = 1}^m {\cos } } \left( {2\pi \Delta f l\xi T} \right)\notag\\
=&- \frac{{N - 1}}{2} + \sum\nolimits_{m = 1}^{N - 1} {\frac{{\sin \left( {\pi \Delta f\left( {2m + 1} \right)\xi T} \right)}}{{2\sin \left( {\pi \Delta f\xi T} \right)}}} \notag\\
=&- \frac{{N - 1}}{2} + \frac{{\sin \left( {\pi \Delta f\left( {N - 1} \right)\xi T} \right)}}{{2\sin \left( {\pi \Delta f\xi T} \right)}}\notag\\
&\left( {\frac{{\sin \left( {\pi \Delta fN\xi T} \right)}}{{\sin \left( {\pi \Delta f\xi T} \right)}}\cos \left( {\pi \Delta f\xi T} \right) \!+\! \cos \left( {\pi \Delta fN\xi T} \right)} \right)\notag\\
=&- \frac{N}{2}+\frac{\sin^2(\pi N \Delta f \xi T)}{2\sin^2 (\pi \Delta f \xi T)}.
\label{Deriving_ISI3}
\end{align}
Plugging~\eqref{Deriving_ISI2} and~\eqref{Deriving_ISI3} back into~\eqref{Deriving_ISI1} results in~\eqref{Accumulated_ISI_final} after some mathematical manipulations.
\hfill $\blacksquare$
%%%%%%%%%%%%%%%%

%%%%%%%%%%%%%%%%
\bibliographystyle{IEEEtran}
\bibliography{reference}
\end{document}